\documentclass[twocolumn,showpacs,preprintnumbers,amsmath,amssymb,nofootinbib]{revtex4}
\usepackage{graphicx}% Include figure files
\usepackage{dcolumn}% Align table columns on decimal point
\usepackage{bm}% bold math

\begin{document}

\preprint{APS/123-QED}

\title{Collisionally Regenerated Dark Matter Structures in Galactic Nuclei}
% Force line breaks with \\

\author{David Merritt}
 \email{merritt@astro.rit.edu}
\author{Stefan Harfst}%
 \email{harfst@astro.rit.edu}
\affiliation{%
Department of Physics, Rochester Institute of Technology,\\
85 Lomb Memorial Drive, Rochester, NY 14623
}%

\author{Gianfranco Bertone}
 \email{bertone@pd.infn.it}
\affiliation{INFN, Sezione di Padova, Via Marzolo 8, Padova, 
I-35131, Italy}%

\date{\today}% It is always \today, today,

\begin{abstract}
We show  that the presence of a $\rho\sim r^{-3/2}$
dark matter overdensity can be robustly
predicted at the center of any galaxy old enough to have grown a power-law
density cusp in the {\it stars} via the Bahcall-Wolf mechanism.
Using both Fokker-Planck and direct $N$-body integrations,
we demonstrate collisional generation of these dark matter
``crests'' (Collisionally REgenerated STtructures) even in the 
extreme case that the density of both stars and
dark matter were previously lowered by slingshot ejection
from a binary supermassive black hole.
The time scale for collisional growth of the crest is approximately
the two-body relaxation time as defined by the stars, which
is $\lesssim 10$ Gyr at the centers
of stellar spheroids with luminosities
$L\lesssim 10^{9.5}L_\odot$, including the bulge of the Milky Way.
The presence of crests can robustly be
predicted in such galaxies, unlike the steeper enhancements, called
``spikes'', produced by the adiabatic growth of black holes.
We discuss special cases where the prospects for detecting dark 
matter annihilations from the centers of galaxy haloes are significantly
affected by the formation of crests.  
\end{abstract}

\pacs{Valid PACS appear here}% PACS, the Physics and Astronomy
\maketitle

\section{\label{sec:intro}Introduction}

While the evidence for a dynamically significant component
of dark matter (DM) on cosmological scales is compelling,
the nature of the DM is unknown, and its distribution
on sub-galactic scales remains uncertain.
A widely discussed DM candidate is the supersymmetric 
neutralino \cite{BHS-04,Bergstrom:2000pn}, the presence of which might be detected
indirectly through the products (gamma rays, neutrinos, anti-matter)
of its self-annihilations \cite{MPLA,Carr-06}.
In the case of photons, the annihilation signal is simply proportional to the square
of the DM density $\rho_\chi$ integrated along the line of sight,
and most discussions of indirect detection have
focussed on the centers of galaxies, including the
Milky Way galaxy, where the DM density is likely to be highest 
\cite{Stecker-88,BUB-98,BSS-01,FPS-04,Hooper-04}.

Supermassive black holes (SBHs) are believed to be
generic components of galactic nuclei \cite{FF-05}
and are expected to strongly 
influence the distribution of mass (stars, DM) 
at distances $\lesssim r_h$ from the SBH \cite{RPP-06}, 
where $r_h$ is the gravitational influence 
radius, defined as the radius within which the gravitational force
from the SBH dominates that from the stars.
In the case of the Milky Way SBH, $r_h\approx 3$ pc.
In one widely discussed model \cite{GS-99}, 
a so-called {\it spike} forms around the SBH
at $r\lesssim r_h$ as it grows adiabatically.
The density in such a spike is a steep function of radius
near the SBH,
$\rho_\chi\sim r^{-\gamma}$, $\gamma\gtrsim 2$
implying a large DM annihilation rate
\cite{GS-99,Bertone:2002je,ABO-04}.
However the formation of such spikes requires finely-tuned
initial conditions \cite{UZK-01} and the spikes
are easily destroyed 
\cite{MMVJ-02,BM-05}.
Furthermore, the stars would react in the same was as the
DM to spherically-symmetric growth of a SBH \cite{RPP-06},
and in the galaxies with long central relaxation times where
steep stellar cusps
%\footnote{In keeping with established
%usage, we refer to a 
%power-law distribution of DM around a SBH as a ``spike,''
%and a power-law distribution of stars around a SBH
%as a ``cusp.''}  
could persist for $10$ Gyr or longer,
none are seen \cite{RPP-06}.
In such galaxies,
stellar density profiles are typically flat at $r\lesssim r_h$,
believed to be a consequence of the ``scouring'' effect
of binary SBHs during galaxy mergers \cite{Merritt-06}.
The density of DM at the centers of these galaxies would also 
presumably be low.

Steeply-rising stellar densities {\it are} instead observed at 
$r\lesssim r_h$ in the bulge of the Milky Way and 
possibly in M32, 
a nearby dwarf elliptical galaxy \cite{Lauer-98,Alexander-99}.
In both of these dense, compact stellar systems, 
the two-body relaxation time near the SBH -- the time for 
stars to exchange orbital
energy via gravitational encounters -- is 
$\lesssim 10^{10}$ yr, short enough for
the formation of a Bahcall-Wolf \cite{BW-76} 
(``collisional'') density cusp in the stars
around the SBH, $\rho_\star\sim r^{-7/4}$.
In dense galaxies like these, a Bahcall-Wolf cusp in the stars
can even re-form after being destroyed by a binary SBH \cite{MS-06}.

In this paper, we discuss the evolution of the DM 
density in a nucleus that grows a collisional cusp 
in the stars via the Bahcall-Wolf mechanism.
The DM particles are essentially collisionless, 
but they scatter off of stars
\cite{IZG-04,Merritt-04,GP-04}, forming 
a $\rho_\chi\sim r^{-3/2}$ density ``crest'' (Collisionally
REgenerated STructure) near the SBH
in roughly one stellar relaxation time.
Remarkably, as we show, this is true even in the case that the 
galaxy  core was previously ``scoured'' by a binary SBH:
DM particles are scattered by stars into regions
of phase space corresponding to tightly-bound orbits
around the SBH that were previously depleted by the binary.

In Sects. II and III we present Fokker-Planck, as well as direct
$N$-body, integrations of a combined, star+DM system around
a central point mass that demonstrate the formation of
crests on roughly a star-star relaxation time scale.
Sect. IV discusses the envirnomental conditions necessary for
the formation of crests; we show that these conditions
are likely to be satisfied in stellar spheroids
comparable in luminosity to that of the Milky Way or fainter,
allowing us to robustly predict the presence of DM crests 
in these systems.
In Sect. V we discuss special cases where the prospects for detecting dark 
matter annihilations from the centers of galaxy haloes are significantly
affected by the formation of crests.
Our conclusions are summarized in Sect. VI.

\begin{figure*}
\includegraphics[angle=-90.,width=0.8\textwidth]{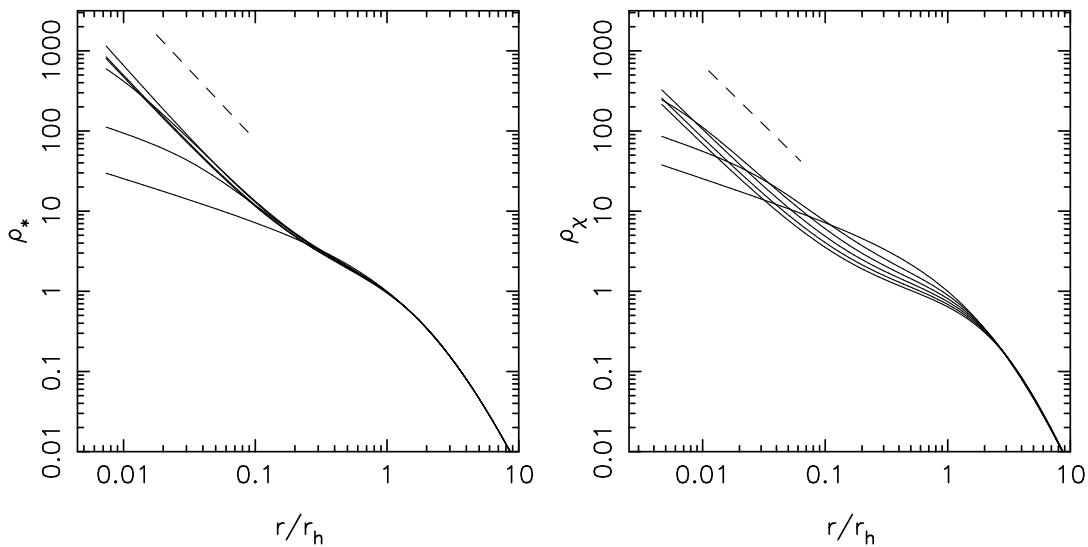}
\caption{\label{fig:fp1} Solutions of the Fokker-Planck
equations~(\ref{eq:fp2a})-(\ref{eq:DEE2})
that describe the joint evolution of stars and dark matter
around a black hole due to star-star and
star-DM gravitational encounters.
Length unit $r_h$ is the radius containing a mass in
stars equal to twice the black hole mass at $t=0$
(roughly 3 pc at the Galactic center).
Density is in units of its initial value at $r_h$.
Curves show the stellar (left) and dark matter (right) density
profiles
at times $(0,0.2,0.4,0.6,0.8,1.0)$ in units of the initial
relaxation time (Eq.~\ref{eq:TR}) at $r=r_h$.
Dashed lines are the ``steady-state'' solutions,
Eqs.~(\ref{eq:sanal}) and~(\ref{eq:ganal}).}
\end{figure*}

\section{\label{sec:FP}Fokker-Planck Treatment}

An approximate description of the combined evolution 
of stars and DM at the center of a galaxy
is provided by the isotropic, multi-mass
Fokker-Planck equation  \cite{Henon-61,Spitzer-87,Merritt-83}.
Let $f(E,m,t)dm$ be the number density in phase space
of objects (stars, DM particles)  in the mass range
$m$ to $m+dm$; $E\equiv -v^2/2+\phi \ge 0$ is the 
binding energy per unit mass
and $\Phi\equiv -\phi$ is the gravitational potential,
assumed fixed in time.
(In what follows, changes in $f$ are only significant 
within the SBH's sphere of influence and the assumption 
of a fixed potential is reasonable.)
Then 
\begin{subequations}
\begin{eqnarray}
{\partial f\over\partial t} &=& {1\over 4\pi^2p}
{\partial \over\partial E} 
\left(m D_Ef + D_{EE}{\partial f\over\partial E}\right),
\label{eq:fp1}\\
D_{E}(E,t) &=& -16\pi^3\Gamma \int_E^\infty dE' p(E') g(E',t), 
\label{eq:DE1}\\
D_{EE}(E,t) &=& -16\pi^3\Gamma  
\bigg[q(E) \int_0^E dE'h(E',t) + \nonumber \\
& & \int_E^\infty dE' q(E')h(E',t)\bigg].
\label{eq:DEE1}
\end{eqnarray}
\end{subequations}
The function $p(E) = 4\sqrt{2}\int_0^{r_{max}(E)} dr r^2 \sqrt{\phi(r)-E}
=-\partial q/\partial E$
is the phase space volume accessible per unit of energy,
$\Gamma=4\pi G^2\ln\Lambda$,
and $\ln\Lambda$ is the Coulomb logarithm.
The functions $g$ and $h$ are moments over mass of $f$:
\begin{subequations}
\begin{eqnarray}
g(E,t) &=& \int_0^\infty f(E,m,t) m dm, \\
h(E,t) &=& \int_0^\infty f(E,m,t) m^2 dm 
\end{eqnarray}
\end{subequations}
e.g. $g$ is the phase-space mass density.
%We ignore for the moment loss terms associated with scattering of stars
%or DM into the BH, or DM self-annihilations;
%these processes are discussed below.

\begin{figure}
\includegraphics[angle=-90.,width=0.5\textwidth]{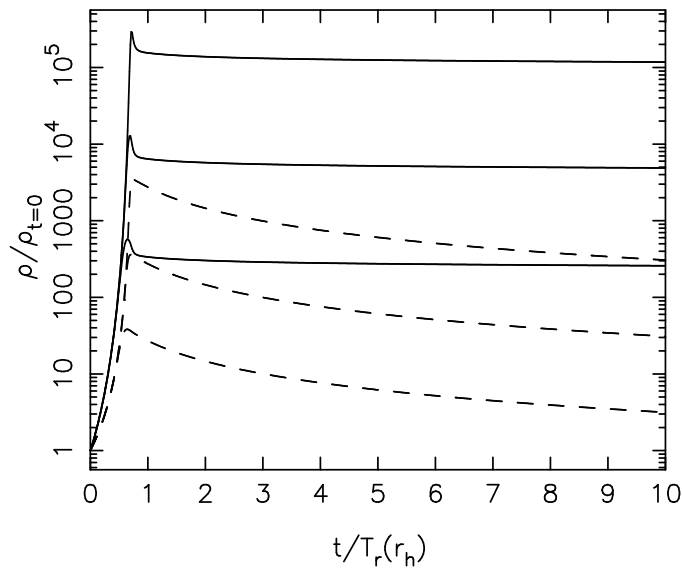}
\caption{\label{fig:fp2} Evolution of the stellar 
(solid lines) and dark matter (dashed line) densities 
at radii of ($10^{-5},10^{-4},10^{-3})r_h$ in the
Fokker-Planck integration of Fig. 1.
Densities are normalized to their values at $t=0$.}
\end{figure}

Consider now a nucleus containing just two components,
stars of mass $m_\star$ and DM particles with
an unspecified range of masses such that
$m_\chi\ll m_\star$.
Assume further that $g_\chi\ll g_\star$.
Taking the first moment of Eq.~(\ref{eq:fp1}) over mass
then yields the two evolution equations
\begin{subequations}
\begin{eqnarray}
{\partial g_\star\over\partial t} &=& {1\over 4\pi^2p}
{\partial \over\partial E} 
\left(m_\star D_Eg_\star + D_{EE}{\partial g_\star\over\partial E}\right),
\label{eq:fp2a}\\
{\partial g_\chi\over\partial t} &=& 
{1\over 4\pi^2p}
{\partial \over\partial E} \left(D_{EE}{\partial g_\chi\over\partial E}\right),
\label{eq:fp2b}
\end{eqnarray}
\end{subequations}
with diffusion coefficients
\begin{subequations}
\begin{eqnarray}
D_{E}(E,t) &=& -16\pi^3\Gamma \int_E^\infty dE' p(E') 
g_\star(E',t),
\label{eq:DE2}\\
D_{EE}(E,t) &=& -16\pi^3\Gamma m_\star 
\bigg[q(E) \int_0^E dE'g_\star(E',t) + \nonumber \\
& & \int_E^\infty dE' q(E')g_\star(E',t)\bigg].
\label{eq:DEE2}
\end{eqnarray}
\end{subequations}
The DM particles, being of negligibly small mass,
do not self-interact gravitationally and they
evolve solely due to heating by (i.e. scattering off of)
the stars \cite{IZG-04,Merritt-04}.
As a result, the characteristic time for change in either
$g_\star$ or $g_\chi$ is the same,
$\left(4\pi\Gamma m_\star g_\star\right)^{-1}$,
equal to within a constant factor to the standard 
two-body relaxation time $T_r$ defined by the stars alone
\cite{Spitzer-87}:
\begin{equation}
T_r \equiv {0.065 v_{rms}^3\over G^2 m_\star \rho_\star \ln\Lambda},
\label{eq:TR}
\end{equation}
with $v_{rms}$ the mean square velocity.
In a time $\gtrsim T_r$, the stellar distribution
approaches its steady-state form near the SBH,
\begin{equation}
g_\star(E)\sim E^{1/4},\ \ \ \ \rho_\star(r)\sim r^{-7/4},
\label{eq:sanal}
\end{equation}
the so-called Bahcall-Wolf \cite{BW-76} solution.
The steady-state solution for the DM is obtained by setting
$\partial g_\chi/\partial E=0$, or 
\begin{equation}
g_\chi(E)\sim {\rm const.},\ \ \ \ \rho_\chi(r)\sim r^{-3/2},
\label{eq:ganal}
\end{equation}
\cite{BW-77,GP-04,Merritt-04}.
Both of these solutions assume a non-evolving phase space
density far from the SBH;
in reality, the DM density will drop after the formation
of the crest, due to ongoing heating by the stars \cite{Merritt-04},
and under certain circumstances the stellar density
may continue to evolve as well \cite{RPP-06}
(the functional form (\ref{eq:ganal}) 
of $g_\chi$ is unaffected by such evolution and we ignore
it in what follows).
While the solution (\ref{eq:ganal}) for $g_\chi$ at $r\ll r_h$
is independent of $g_\star$, the time
required to attain this density profile, and the subsequent
rate of heating by the stars, do depend on $g_\star$.

\begin{figure*}
\includegraphics[angle=-90.,width=0.8\textwidth]{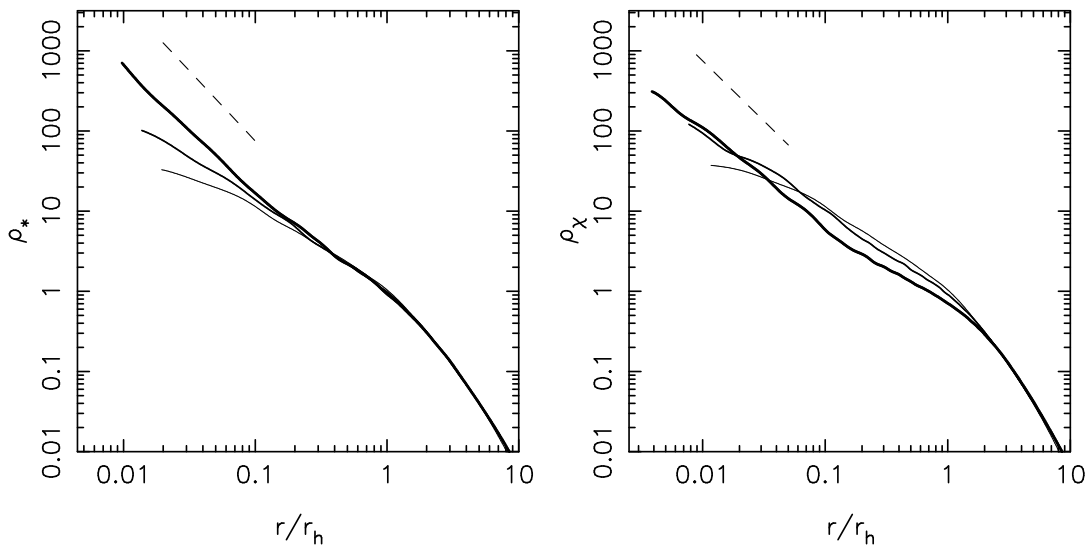}
\caption{\label{fig:NB} Direct $N$-body integration of
``star'' (left) and ``dark matter'' (right) particles 
around a single massive particle (``black hole''), 
starting from a galaxy model in which a pre-existing
binary black hole had created a low-density core
\cite{MS-06}.
Thin, normal and thick curves show the density profiles at
times ($0,0.25,1.0$) in units of $T_r(r_h)$.
Dashed lines are as in Fig.~1.}
\end{figure*}

Fig.~\ref{fig:fp1} shows time-dependent solutions to Eqs.~(\ref{eq:fp2a})-
(\ref{eq:DEE2}).
Initial density profiles for both the stars and DM
were assigned as in Ref.~\cite{Dehnen-93};
these models have 
$\rho\propto r^{-\gamma}$ near the SBH, and
we chose $\gamma=0.5$, the flattest density profile 
consistent with an isotropic phase-space distribution in a $1/r$ potential.
(The density at large radius in this model falls off more steeply
than  the $\rho_\chi\sim r^{-1}$ dependence predicted
in standard DM halo models; however we are concerned here
with the evolution only at very small radii.)
The unit of time in Fig.~\ref{fig:fp1} is the relaxation time
for the stars at the influence radius $r_h$, defined
in the standard way as the radius containing a mass in stars 
equal to twice $M_\bullet$.
In principle, $r_h$ so defined changes with time in response
to the changing stellar density, but this effect is almost
negligible and we ignore it in what follows.

In a time $\sim 0.5 T_r(r_h)$, the stars are seen to attain the 
Bahcall-Wolf profile at $r\lesssim 0.2r_h$.
The evolution of the DM is more complex.
A $\rho_\chi\sim r^{-3/2}$ crest inside $r\approx 0.1 r_h$
is formed in roughly the same time.
However the amplitude of the crest
drops thereafter as the DM is heated by the stars
(Figs.~\ref{fig:fp1},\ref{fig:fp2}):
by a factor $1/e$ in a time of $\sim 1.2 T_r(r_h)$ after
peak density, and 
by a factor $1/e^2$ in a time of $\sim 4.5 T_r(r_h)$.
Evolution of the DM density profile after crest
formation is approximately self-similar,
$\rho_\chi(r,t)\approx \rho_{\chi,0}(r)G(t/T_r)$ 
at $r\lesssim r_h$, with $dG/dt<0$.

\section{\label{sec:NB}$N$-Body Treatment}

The Fokker-Planck equations (\ref{eq:fp2a})-(\ref{eq:DEE2}) embody
a number of approximations
(isotropy, small-angle scattering, uncorrelated encounters, 
fixed gravitational potential, etc.)
that may be violated in real stellar systems.
Furthermore the initial conditions of Fig.~\ref{fig:fp1}
are ad hoc.
We therefore carried out a direct $N$-body integration
of stars and massless particles around a point mass, 
starting from initial conditions
that realistically represent the center of a galaxy after
a binary SBH has created a low-density core \cite{MS-06}.

The $N$-body code was adapted from $\varphi$GRAPE \cite{Harfst-06},
a direct-summation code that advances particles via a fourth-order
integrator and computes gravitational forces via calls to 
special-purpose accelerator boards called GRAPEs \cite{GRAPE}.
The code was modified to run in serial mode using a single GRAPE-6
computer which has an onboard memory limit of $\sim 256K$ particles
and a speed of $\sim 1$ Tflops.
The code was also modified to include massless (DM) particles;
since these do not ``see'' each other gravitationally,
they need not all be loaded into GRAPE memory simultaneously,
allowing the GRAPE's memory limit to be circumvented.
We used $N_\star=1.2\times 10^5$ ``star'' particles
and $N_\chi=2.4\times 10^6$ ``DM'' particles.

Initial conditions for the $N$-body integration were adapted
from Run 8 of Ref.~\cite{MS-06}.
In that paper, $N$-body simulations were used to follow the
formation of low-density cores via inspiral of a SBH into
the  center of a galaxy containing a second, more massive SBH;
in Run 8, the binary mass ratio was $1:4$.
We replaced the two massive particles at the final time
step of Run 8 by a single particle (the SBH) having their
combined mass ($0.0125$ in units of the total galaxy mass),
and positioned this particle at the center of 
mass of the pre-existing binary.
Massless (DM) particles were then added via a bootstrap algorithm:
a star particle was chosen at random; a DM particle
was placed randomly on a sphere with radius equal to that of
the star particle; the DM particle was assigned a velocity
with random direction subject to the constraint that
the radial and tangential components were equal to those
of  the star particle.
This scheme produced a model in which the DM had the
same initial phase-space distribution as the stars; in particular,
both components had essentially flat central density profiles
(Fig.~\ref{fig:NB}).
In addition, the velocity distributions at $r\lesssim r_h$
were biased toward
circular motions, a result of gravitational slingshot ejection 
of particles on radial orbits by the binary.
$N$-body integration of this model until a time $\sim T_r(r_h)$
using $\varphi$GRAPE required $\sim 43$ d on the GRAPE-6
special-purpose computer, for a total of 
$\sim 6.6\times 10^{17}$ floating-point operations.

Fig.~\ref{fig:NB} shows the results.
The star particles form a Bahcall-Wolf cusp in a time
$\sim 0.5 T_r(r_h)$.
The response of the DM particles is also consistent
with what was found in the Fokker-Planck integration 
(Fig.~\ref{fig:fp1}):
formation of a crest
at $r\lesssim 0.1r_h$ and a gradual drop in $\rho_\chi$
as the DM is heated by the stars.
The DM density slope at $r\lesssim 0.01 r_h$ is not
well constrained due to the finite number of particles
but is consistent with $\rho_\chi\propto r^{-3/2}$.
Henceforth we will assume that the Fokker-Planck solution
correctly describes the radial dependence of the DM
density at radii too small to be resolved via the
$N$-body integration.

Both the size of the core formed by an inspiralling SBH
(measured in units of the binary's mass)
and the time scale to regrow a stellar cusp 
(measured in units of $T_r(r_h)$) are
almost independent of the binary mass ratio 
\cite{MS-06,Merritt-06}
and so the results of our single $N$-body 
integration should be representative of the majority of
galactic nuclei that formed via mergers,
if the length and mass are scaled to $r_h$ and $M_\bullet$
respectively.

\section{\label{sec:gal}Galaxy Properties Relevant to 
the Formation of Crests}

Several basic conditions must be satisfied for a 
DM crest to form in a galactic nucleus.
(1) The two-body (star-star) relaxation time must be short.
(2) The nucleus must contain a massive BH.
(3) For the annihilation signal to be detectable,
there must be a significant amount of DM 
on scales $\lesssim r_h$.
In this section, we review current knowledge
concerning physical conditions in galactic nuclei
and discuss which kinds of galaxies are most likely
to harbor DM crests.

\begin{figure}
\includegraphics[width=0.425\textwidth]{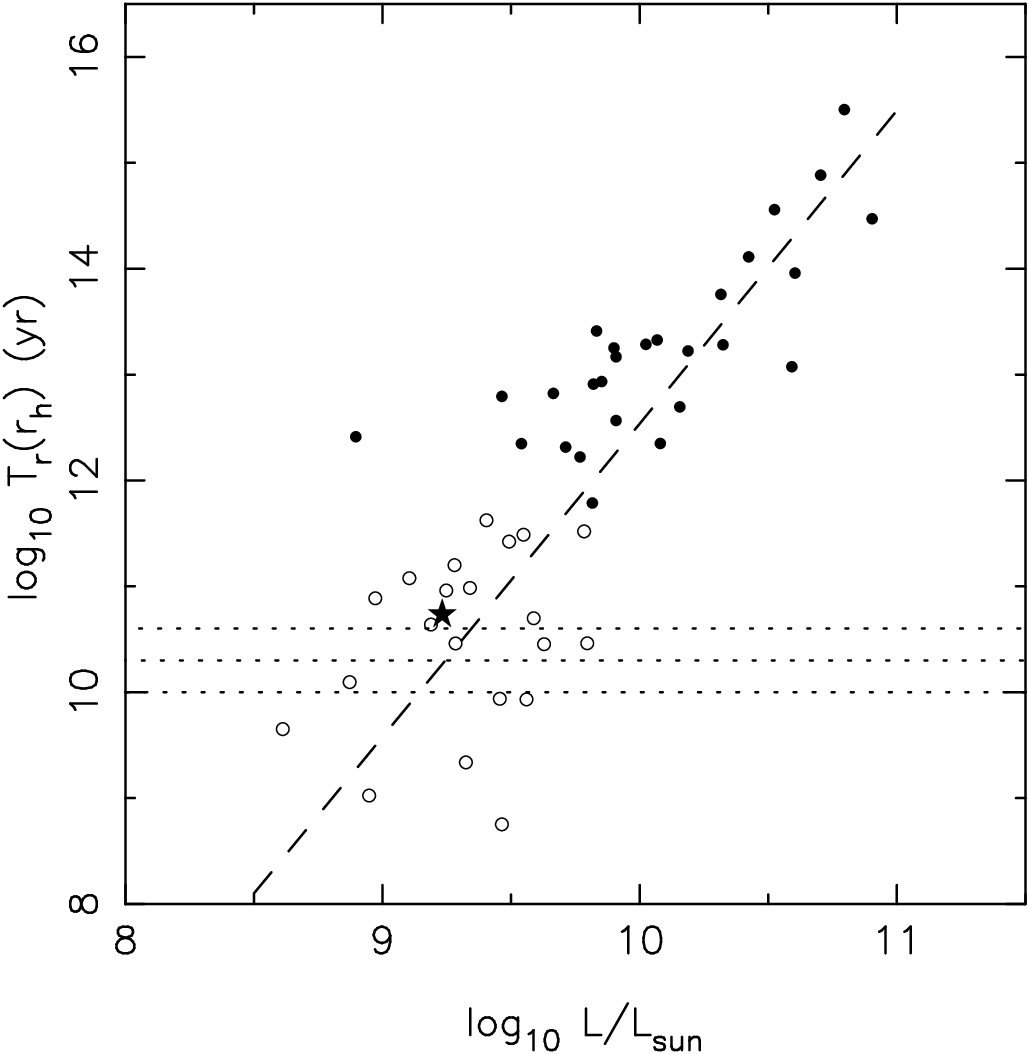}
\caption{\label{fig:tr} Relaxation times at the SBH's
influence radius $r_h$ in Virgo cluster 
galaxies (circles) and the Milky Way (star)
vs. the total blue luminosity of the galaxy
(in the case of the Milky Way, the bulge).
A  stellar mass of $1M_\odot$ was assumed when computing $T_r$.
Filled circles: Virgo cluster galaxies in which $r_h$ is resolved, i.e.
$r_h\ge 0.1''$.
Open circles: Virgo cluster galaxies in which $r_h$ is unresolved.
Stellar luminosity profiles were taken from Ref. \cite{ACS6}
and black hole masses were computed from the $M_\bullet-\sigma$
relation \cite{FF-05}, except in the case of the Milky Way 
for which $M_\bullet$ has been directly determined from
stellar orbits \cite{Ghez-05}.
Horizontal dotted lines are at $T_r(r_h)=(1,2,4)\times 10^{10}$ yr.
Dashed line is a regression fit to the data,
Eq.~\ref{eq:tr}.
}
\end{figure}

\subsection{Relaxation times}
The two-component models presented above
demonstrate that a stellar cusp and its associated DM
crest are generated in a time of $\sim 0.5T_r(r_h)$.
(The single-component $N$-body studies in 
Refs.~\cite{Preto-04,Baumgardt-04,MS-06} reach
similar conclusions about the time scale for stellar
cusp formation.) 
Fig.~\ref{fig:tr} shows estimates of $T_r(r_h)$ 
vs. spheroid luminosity in a complete sample of early-type
(elliptical and lenticular) galaxies in the Virgo
cluster \cite{ACS6,ACS8} and the bulge of the Milky Way.
(The spheroid is defined as the entire luminous galaxy
in the case of elliptical and lenticular galaxies,
and as the bulge component in the case of spiral galaxies.)
There is a well-defined trend of $T_r(r_h)$ with spheroid
luminosity $L$:
\begin{equation}
T_r(r_h)\approx 3.8\times 10^9 {\rm yr}\ L_9^{3.0}
\label{eq:tr}
\end{equation}
where $L_9\equiv L/10^9 L_\odot$.
Spheroids with $L\lesssim 2\times 10^9 L_\odot$ have
$T_r(r_h)\lesssim 3\times 10^{10}$ yr, which is short enough
for the formation of a crest, assuming that the SBH and spheroid
were in place at least $\sim 10^{10}$ yr ago.
The Milky Way (marginally) satisfies this condition,
and the Galactic center stellar cluster in fact has a density profile
that is consistent the Bahcall-Wolf form \cite{Alexander-99,Lauer-98}.

Conditions for the formation of crests are relaxed somewhat
if there is a top-heavy spectrum of stellar masses since
the DM scattering time scales as $\tilde m_\star^{-1}$ where
$\tilde m_\star = \langle m_\star^2\rangle/\langle m_\star\rangle$
\cite{Merritt-04}.
The stellar cusp can also evolve more quickly in this case
\cite{Baumgardt-04b}.

Unfortunately, in the luminosity range most relevant
to DM crest formation, the Milky Way bulge is the
{\it only} spheroid near enough for a 
Bahcall-Wolf cusp to be detected even if present.
Spheroids fainter than $L\approx 10^9L_\odot$ and outside
the Local Group are always unresolved on scales $r\lesssim r_h$
(e.g. Fig.~\ref{fig:tr}).
However there are indications that nuclear
structure  begins to change systematically as
$L$ drops below $\sim 10^9L_\odot$, since an 
increasingly large fraction of spheroids exhbit
compact stellar nuclei \cite{Boeker-04,ACS8}.
Only one compact nucleus is spatially well resolved, 
in the Local Group galaxy NGC 205 ($L\approx 10^{8.4}L_\odot$), 
and the relaxation time is found to drop to $\sim 10^8$ yr
at the center \cite{N205}; 
this value is derived assuming no SBH since NGC 205
shows no evidence of rising velocities near the center
\cite{N205}.
Whether the compact nuclei in faint spheroids
are consequences of their
short nuclear relaxation times \cite{HHK-91}, 
the absence of SBHs in these galaxies \cite{ACS-nuclei,Wehner-06},
gas-dynamical processes \cite{King-06}, 
or some combination of these factors is currently unclear.
However the form of a DM crest is essentially independent
of $\rho_\star(r)$ (\S 2) and if the compact nuclei satisfy
the other two conditions identified above
(presence of a SBH and a significant amount of DM;
see discussion below) they
would be expected to contain DM crests as well.

If a nucleus is much older than $T_r(r_h)$,
a DM crest will form
then decay in amplitude as the DM particles attempt to
reach equipartition with the much heavier stars
(Fig.~\ref{fig:fp1}; \cite{Merritt-04}).
Fig.~\ref{fig:fp2} suggests that the 
crest density drops by a factor $10$ from its peak
value -- corresponding to two orders
of magnitude in the annihilation signal -- 
in a time $\sim 8-9 T_r(r_h)$.
The trends in Fig.~4 are too poorly defined at low
values of $L/L_\odot$ to allow us to state clearly
for which galaxies this would occur, but a straightforward
extrapolation of Eq.~(\ref{eq:tr}) suggests that
$T_r(r_h)$ falls below $\sim 10^9$ yr at 
$L\approx 10^{8.7}L_\odot$.
If this inference is correct, it follows that crests 
would be present, with reasonable amplitudes, only for a 
fairly narrow range of spheroid properties: i.e. spheroids
older than $\sim$one relaxation time but younger than many relaxation times.
Taking into account the scatter in  
Fig.~\ref{fig:tr}, the corresponding range in spheroid
luminosities might be approximately
$3\times 10^8 L_\odot\lesssim L\lesssim 3\times 10^9 L_\odot$.

If low-luminosity spheroids do {\it not} contain massive BHs,
nuclear relaxation times might generically be as short as those observed
in NGC 205 and M33, i.e. $10^7-10^8$ yr.
In this case, the stars would undergo core collapse in 
$\lesssim 10^{10}$ yr producing a $r^{-2.25}$ density profile,
and the DM density would be expected to evolve only 
slightly \cite{Kim-04}.

At the other extreme of luminosity,
galaxies with $L\gtrsim 10^{10}L_\odot$ are always 
observed to have low-density cores with radii
$\sim r_h$ \cite{MM-02,Ravin-02,Graham-04},
and high central velocity dispersions,
implying long nuclear relaxation times
(Fig.~\ref{fig:tr}).
In principle, these galaxies could still harbor 
{\it collisionless} DM spikes around their SBHs \cite{GS-99}, 
but this seems unlikely:
the nuclear structure of these galaxies is consistent
with no dynamical evolution having occurred since
the most recent merger ``carved out'' the
luminous core \cite{Merritt-04}, and the same merger 
event that created the stellar core would have destroyed a
density spike in the DM \cite{MMVJ-02}.
Below we discuss the possibility that {\it low}-luminosity
spheroids might partially retain their steep,
collisionless spikes; these galaxies are unlikely
to have experienced significant mergers and scattering off
of stars need not completely convert such a spike into
a weaker, Bahcall-Wolf crest in a Hubble time \cite{Merritt-04}.

\subsection{Black holes}

The presence of a massive BH is another necessary condition 
for the formation of a DM crest.
Robust detection of SBHs is possible only in galaxies near 
enough that the stellar or gas kinematics can be resolved
on scales $\ll r_h$.
Reliable, dynamical SBH masses have been determined in a handful of 
such galaxies \cite{FF-05}, and the large-scale properties 
(mass, luminosity, velocity dispersion) of their host
spheroids are found to obey tight scaling relations with $M_\bullet$
\cite{FM-00,Graham-01,MH-03}, e.g.
\begin{equation}
M_{\bullet,8} = (1.66\pm 0.24) \sigma_{200}^{4.86\pm 0.43}
\label{eq:msigma}
\end{equation}
\cite{FF-05}, where $M_{\bullet,8}=M_\bullet/10^8M_\odot$
and $\sigma_{200}$ is the 1D stellar velocity dispersion
in units of $200$ km s$^{-1}$. 
%$10^9 L_\odot \lesssim L \lesssim 10^{11} L_\odot$.
%($10^{6.5}\lesssim M_\odot \lesssim 10^{9.5}$)
The existence of these tight relations suggests that SBHs are 
ubiquitous in bright spheroids,
but the faintest galaxy for which $M_\bullet$
has been robustly determined is M32 
($L\approx 4\times 10^8 L_\odot$; in fact this galaxy
is believed to be the remnant core of a once much
brighter galaxy)
and it is dangerous to assume that relations like 
(\ref{eq:msigma}) apply to fainter systems, 
including the spheroids with $L\lesssim 10^9L_\odot$ that
are most relevant to formation of crests (Fig.~\ref{fig:tr}).

Evidence for massive BHs in spheroids fainter than
$\sim 10^9L_\odot$ -- dwarf elliptical galaxies and
the bulges of late-type spiral galaxies -- comes
almost entirely from the subset of nuclei
that are ``active,'' i.e. that emit a significant
fraction of their energy non-thermally \cite{Ho-04}.
$M_\bullet$ in these galaxies is determined indirectly
by applying empirical relations established in more luminous
active galaxies, e.g. between the size of the so-called
broad emission line region and the nuclear continuum luminosity
\cite{Kaspi-04}.
The presence of BHs with masses as low as $\sim 10^5M_\odot$
has been inferred \cite{Greene-04}; with one exception
(POX 52, a dwarf elliptical galaxy \cite{Barth-04}),
all of the host spheroids are spiral-galaxy bulges.
The BH masses in these galaxies appear to be consistent
with relations like (\ref{eq:msigma}) \cite{Barth-05}.
%There are indications that $L$ is not a good predictor
%of $M_\bullet$ in these systems; for instance, 
%NGC 4395 is a spiral galaxy with $M_\bullet\approx 10^5 M_\odot$ 
%but which lacks a significant bulge component \cite{FH-03}.

Whether BHs with $M_\bullet\lesssim 10^6M_\odot$, 
are present in all low-luminosity spheroids
is still unclear.
In the Local Group, neither the dwarf elliptical NGC 205 
($L\approx 10^{8.4}L_\odot$) or M33 (an apparently
spheroid-less spiral galaxy) appear to have central
BHs although the upper limits on $M_\bullet$ in these
galaxies (based on stellar kinematics) are only marginally 
inconsistent with the scaling relations established
in brighter galaxies \cite{N205,M33}.
A significant fraction ($\sim 40\%$) of nearby galaxies show
evidence of low-level nuclear activity but the 
activity need not be driven by BH accretion in every
case \cite{Ho-04}.
Even if small BHs were present at one time in all low-luminosity
spheroids, a number of processes are capable
of ejecting them from such environments \cite{MMFHH-04}.

Some nuclei might contain binary or multiple BHs,
if the BHs that were deposited there during a galaxy
merger failed to coalesce.
In the presence of a binary SBH, growth of a collisional
cusp  would be inhibited as the binary continued to eject stars
and DM particles from the nucleus via the gravitational slingshot
\cite{MV-92}.
Time scales for binary SBH coalescence due
to gravitational wave emission are longer
than $10^{10}$ yr unless the binary separation $a$
drops below 
\begin{equation}
a_{GW}\approx 2\times 10^{-3}{\rm pc} {q\over (1+q)^2}
M_{12,6}^3
\end{equation} 
where $q=M_2/M_1\le 1$ is the binary mass ratio
and $M_{12,6}=(M_1+M_2)/10^6M_\odot$ \cite{Living}.
This separation is $1-2$ orders of magnitude less
than the separation $a_h$ at which the two BHs become
gravitationally bound, and it is possible for the
binary to stall at $a\approx a_h\gg a_{GW}$ 
(the ``final parsec problem'').
Stalling is least likely in nuclei with
short relaxation times, since stars will scatter
into the binary's sphere of influence where
they can extract angular momentum from the binary
\cite{Yu-02,MM-03}.
A number of other mechanisms have
been identified that can accelerate the evolution
of binary SBHs, even
in collisionless nuclei \cite{Berczik-06,MM-06}.
Constraints on the binarity of the MW SBH are
fairly  tight \cite{Living}, and only
one clear detection of a binary SBH has so far 
been made \cite{Rodriguez-06}, suggesting that they may
be rare.

\subsection{Dark matter}

Detectability of DM crests depends critically on
the DM density at $r_h$, roughly the outer boundary
of the region in which the mass distribution is
modified by the (single or binary) SBH.
Traditionally there have been two approaches
to estimating $\rho_\chi$ at the centers of galaxies;
unfortunately they lead to rather different conclusions
about $\rho_\chi(r_h)$.

{\sl $N$-body simulations of gravitational clustering}
follow the growth of DM halos as they evolve 
via mergers in an expanding, cold-dark-matter ($\Lambda$CDM) universe.
Halo density profiles in these simulations are well determined 
on scales $10^{-2}\lesssim r/r_{vir} \lesssim 10^0$,
where the virial radius $r_{vir}$ is of order
$10^2$ kpc for a galaxy like the Milky Way; 
hence inferences
about $\rho_\chi$ on scales of $r_h\approx 10^{-3}$ kpc
require a radical extrapolation from the $N$-body results.
A standard parametrization of $\rho_\chi$ in these
simulated halos is
\begin{equation}
\rho_\chi(r) = \rho_0\xi^{-1}\left(1+\xi\right)^{-2}
\label{eq:NFW}
\end{equation}
\cite{NFW-96}, the ``NFW profile,''
where $\xi=r/r_s$ and $r_s$ is a scale length
of order $r_{vir}$.
In the Milky Way, $r_{vir}\gg R_\odot$ (the radius of the
Solar circle) hence Eq.~(\ref{eq:NFW}) is essentially a power law 
at $r<R_\odot$ and the implied DM density at $r_h$ is
\begin{equation}
\rho_\chi(r_h)\approx 30 M_\odot {\rm pc}^{-3} 
\left({\rho_\odot\over 10^{-2}M_\odot {\rm pc}^{-3}}\right)
\left({R_\odot\over 8\ {\rm kpc}}\right)
\left({r_h\over 3\ {\rm pc}}\right)^{-1}
\label{eq:NFW2}
\end{equation}
where $\rho_\odot\equiv\rho_\chi(R_\odot)$ and
$\rho_\odot\approx 8\times 10^{-3}M_\odot {\rm pc}^{-3}$
(from the Galactic rotation curve).
Moore et al. \cite{Moore-98} argue for a steeper inner slope,
$\rho_\chi\sim r^{-1.5}$, implying 
$\rho_\chi(r_h)\approx 10^3 M_\odot {\rm pc}^{-3}$.
Still higher DM densities could exist if 
the baryons (stars, gas) lose energy radiatively and contract, 
deepening the potential well and pulling in the DM
(e.g. \cite{Prada-04}).
Based on halo scaling relations 
\cite{Empirical3},
DM densities at $r_h$ would be higher in spheroids fainter
than that of the Milky Way.

{\sl Rotation-curve studies of dark-matter-dominated galaxies}
are generally interpreted as implying much lower, central
DM densities
\cite{Burkert-95,SB-00,Blok-02,Gentile-05,Blok-05}.
While there are caveats to this interpretation --
systematic biases in long-slit observations \cite{SGH-05},
non-circular motions \cite{Simon-05},
gas pressure \cite{valenzuela-05}, etc. --
these effects do not seem capable of fully explaining  the
discrepancies between rotation curve data and 
expressions like (\ref{eq:NFW}) \cite{Blok-04,Gentile-05}.
A model for $\rho_\chi(r)$ that is often fit to rotation
curve data is \cite{Burkert-95},
\begin{equation}
\rho_\chi(r) = \rho_c\left(1+\xi\right)^{-1}\left(1+\xi^2\right)^{-1},
\label{eq:Burkert}
\end{equation}
the ``Burkert profile,'', where $\xi\equiv r/r_c$ and
$r_c$ is the core radius.
Inferred core radii are $\sim 10^3$ pc and inferred
central densities typically lie in the range
$\rho_c\approx (1-5)\times 10^{-2} M_\odot {\rm pc}^{-3}$.

Several resolutions have been suggested for this apparent conflict
between theory and observation \cite{Tasitsiomi-03}.
Since the $N$-body halos are not resolved on the scales
($\sim 10^2$ pc) where rotation curves are typically measured,
the mismatch may simply be due to a poor choice of empirical
model used to parametrize $\rho_\chi(r)$.
For instance, the alternative parametrization
\begin{equation}
\rho_\chi(r) = \rho'\xi^{-p}\exp\left(-b\xi^{1/n}\right),
\label{eq:PS}
\end{equation}
the ``Prugniel-Simien'' law \cite{PS-87},
is both a better fit to the $N$-body data than Eq.~(\ref{eq:NFW})
and is also in reasonable accord with rotation curve data
\cite{Navarro-04,MNLJ-05,Empirical1,Empirical3}.
Here $\xi\equiv r/R_e$ and $R_e$ is the radius containing
$1/2$ of the projected halo mass 
(the relation between $R_e$ and the virial radius
is discussed in \cite{Empirical2}).
%The dependence of $R_e$ on halo mass is 
%roughly $R_e\propto M_{halo}^{1/2}$
%\cite{Empirical3}.
The profile shape in Eq.~(\ref{eq:PS}) is determined by the curvature parameter
$n$; for the $N$-body halos, $n$ is found
to be a weak function of halo mass, but exhibits a
substantial scatter at all halo masses,
\begin{equation}
2\lesssim n\lesssim 6
\end{equation}
(e.g. \cite{Empirical2}, Fig. 1a).
The constants $b$ and $p$ in Eq.~(\ref{eq:PS}) are
determined uniquely by $n$ \cite{Empirical1};
for the range in $n$ cited above, 
$0.7\lesssim p \lesssim 0.9$, i.e. the density
increases more slowly toward the center than
the NFW \cite{NFW-96} ($\propto r^{-1}$) profile.

%\begin{equation}
%\rho_\chi(r) = \rho_e\exp\left[-C\left(r/r_e\right)^{1/n}\right],
%\label{eq:einasto}
%\end{equation}
%here $n\approx 5$ for galaxy-sized halos and $C$ is defined such 
%that $r_e$ is the radius containing $1/2$ of the total mass.
%The density at $r_h$ implied by a parametrizations like 
%(\ref{eq:einasto}) is $\sim 10\%$ that implied
%by Eq.~(\ref{eq:NFW}), in a galaxy like the Milky Way.
%Alternatively, central halo densities might be lowered 
%due to gravitational heating from the
%baryons \cite{MMDM-02,GZ-02,WK-02}.

\subsection {Summary of Galaxy Properties Relevant to the Formation of Crests}
Relaxation times are short enough for the formation of DM crests
in stellar spheroids with $L\approx 10^{9.5} L_\odot$
or fainter.
However the nuclear structure of faint galaxies
is typically unresolved and relaxation times in
spheroids with $L\lesssim 10^9L_\odot$  are uncertain.
SBHs appear to be ubiquitous in stellar spheroids
brighter than $\sim 10^9L_\odot$, with masses that are
well predicted by the properties of the stellar spheroid
via empirical scaling relations.
A handful of massive ($\sim 10^5 M_\odot$) 
BHs have been detected in fainter 
galaxies via their non-thermal spectral features, 
but it is not clear what fraction of low-luminosity 
spheroids contain BHs.
Dark matter densities at $r\approx r_h$ in spheroids
with $L\approx 10^9 L_\odot$ may be as high as
$10^2-10^3 M_\odot {\rm pc}^2$, if standard 
parametrizations of $\Lambda$CDM halo models are correct;
or as low as
$\sim 10^{-2}-10^{-1}M_\odot {\rm pc}^{-3}$ if
rotation curve studies are to be believed.

\section{\label{sec:obs}Observability of the Crests}

In this section, we consider the implications of
collisionally-generated
DM crests for the rate of particle self-annihilations 
and for the
detectability of the resultant gamma rays.
We consider separately the case of the Milky Way
and external galaxies.
Since the $\rho_\chi\sim r^{-3/2}$ crests considered
here are relatively weak (e.g. compared with the
steeper spikes that form via adiabatic growth of
a SBH, $\rho_\chi\sim r^{-\gamma}$, $2\lesssim\gamma\lesssim 3$ \cite{GS-99}), 
we focus on the question of whether the
DM distribution inferred above at $r\lesssim r_h$
implies a substantial {\it increase} in the predicted
annihilation signal compared with the signal from a 
galaxy that lacks such a crest.

The annihilation rate from neutralinos near the center 
of a spherically-symmetric
DM halo is proportional to $\int\rho_\chi^2(r) r^2 dr$.
In the presence of a $\rho_\chi\sim r^{-3/2}$ crest, 
the integral diverges as $\log(r_0^{-1})$ where
$r_0$ is the inner radius of the crest.
Roughly, $r_0$ is the maximum of $(r_S,r_a)$ 
where $r_S= 2GM_\bullet/c^2$
is the Schwarzschild radius of the SBH and $r_a$ is the
radius where the self-annihilation time equals $\sim T_r$.
For all reasonable values of $m_\chi$ and $\sigma v$,
the annihilation cross section,
$r_a\ll r_S$ hence we assume $r_0\approx r_S$ in what follows.

Evaluating this integral in the case of the $N$-body 
density profile of Fig. 3 at $t=T_r(r_h)$, 
we find
\begin{equation}
\int_{r_0}^\infty\rho_\chi^2(r) r^2 dr = \rho_h^2r_h^3\left[
C_1\ln\left({r_h\over r_0}\right) + C_2\right]
\label{eq:int}
\end{equation}
with $C_1\approx 4.22\times 10^{-2}$, 
$C_2\approx 2.94$ and $\rho_h\equiv \rho_\chi(r_h)$.
(We assumed $\rho_\chi(r) = \rho_1(r/r_1)^{-3/2}$ 
at $r\le r_1=0.01r_h$ where the $N$-body density
is poorly defined, as justified above.)
Roughly 90\% of the integral
comes from matter at $r\le r_h$.
%, hence Eq.~(\ref{eq:int}) 
%essentially measures the contribution from
%DM within the region where its distribution is
%affected by the SBH.

While the DM density is affected by the
SBH at all $r\lesssim r_h$,
the $\rho_\chi\sim r^{-3/2}$ crests 
only extend out to $r\lesssim 0.1 r_h$.
We find that $\sim 20\%$ of the integral (\ref{eq:int})
comes from the crest proper, i.e. from $r\lesssim 0.1r_h$.

Setting $\ln(r_h/r_0)\approx \ln(c^2/v_{rms}^2)\approx 15$,
the integral in Eq.~(\ref{eq:int}) is $\sim 3.6\rho_h^2r_h^3$.
For comparison, the value corresponding to a constant
density within $r_h$
is $\sim 0.33\rho_h^2r_h^3$.

Henceforth we assume that Fig.~3 correctly represents 
the DM distribution at $r\lesssim r_h$ in nuclei that are
$\sim T_r(r_h)$ old, i.e. that
\begin{equation}
\int_{r_0}^{r_h}\rho_\chi^2(r) r^2 dr \approx 3 \rho_h^2r_h^3.
\label{eq:intb}
\end{equation}
The observable annihilation signal will typically include
a contribution from $r\le r_h$, as well as a contribution
from DM beyond $r_h$ that lies within the detector's 
window.
We will consider several possible forms for the large-radius
dependence of $\rho_\chi$ on $r$.

At times later than $\sim T_r(r_h)$, the amplitude of the
crest drops due to continued heating from the stars
(e.g. Fig.~\ref{fig:fp2}).
We discuss the  consequences of this effect for the
observability of crests in more detail below.

\subsection{Milky Way}

The photon flux from neutralino annihilations in the Galactic
halo, observed by a detector with angular acceptance $\Delta\Omega$, is
\begin{subequations}
\begin{eqnarray}
\Phi(E) &=& {1\over 2}{\sigma v\over m_\chi^2} {dN\over dE}
I(\Delta\Omega), \\
I(\Delta\Omega) &=& {1\over 4\pi}
\int_{\Delta\Omega}d\Omega'\int\rho_\chi^2(l) dl. 
\label{eq:flux}
\end{eqnarray}
\end{subequations}
Here $\sigma v$ is the annihilation cross section times relative
velocity (in the nonrelativistic limit), $dN/dE$ is the gamma ray
spectrum per annihilation, and $l$ is the line-of-sight distance.
In the case of telescope centered on  the Milky Way SBH,
the integral can be written
\begin{eqnarray}
I(\Psi) &\approx& {1\over R_\odot^2}
\bigg [ \int_{r_0}^{r_h} \rho_\chi^2(r)r^2dr  \nonumber \\
& + & \int_{r_h}^{\Psi R_\odot} \rho_\chi^2(r) r^2 dr \nonumber \\
& + & \int_{\Psi R_\odot}^{R_\odot} \rho^2_\chi
\left(r^2 - r\sqrt{r^2-\Psi^2 R_\odot^2}\right)\bigg ]
\label{eq:terms}
\end{eqnarray}
where $\Psi^2\equiv \Delta\Omega/\pi$; the inequality
reflects the omission of the contribution from DM outside
the Solar circle, and we have also implicitly assumed
$\Psi R_\odot > r_h$, valid for all current and planned detectors.
(Atmospheric Cerenkov telescopes like HESS,
and the proposed satellite observatory GLAST,
have $\Delta\Omega\approx 5\times 10^{-5}$ sr.)
We equate the first term on the RHS of Eq.~(\ref{eq:terms}) with
Eq.~(\ref{eq:intb}).
%Note that this term contains both the spike,
%as well as the DM at $r_{sp}\lesssim r\lesssim r_h$;
%$r_h$ is roughly the outer limit of the region affected
%by the binary SBH.
To  evaluate the additional terms we require $\rho_\chi(r)$
at $r>r_h$.
A simple model is a power-law, 
\begin{equation}
\rho_\chi(r) = \rho_\chi(r_h)\left(r/r_h\right)^{-\gamma}, \ \ 
r>r_h.
\end{equation}
For $\gamma=1$ this is a reasonable approximation to both
an NFW profile, Eq.~(\ref{eq:NFW}), 
and a Prugniel-Simien profile, Eq.~(\ref{eq:PS}),
at $r<R_\odot$.
Setting $\gamma=1$ gives
\begin{equation}
I(\Psi) \approx \rho^2_\chi(r_h)r_h^3 R_\odot^{-2} \left(3 + 
{\pi\over 2} {\Psi R_\odot\over r_h} - 1 \right)
\end{equation}
where the first term is the contribution from $r\le r_h$
(Fig.~\ref{fig:NB}).
For $\Delta\Omega=5\times 10^{-5}$
($\Psi \approx 4\times 10^{-3}$ sr), roughly 20\% of
$I$ is due to matter at $r\le r_h$ and roughly
$4\%$ from the crest proper.
This result suggests that the detectability of DM self-annihilations
at the Galactic center is not likely to be significantly 
affected by the presence of a crest.
We note that in Ref.~\cite{GP-04}, the contribution of a DM
crest at the Galactic center to the self-annihilation
signal was found to be much greater. The discrepancy arises from the fact
that these authors ignored the third (dominant) term on the right hand side
of Eq.~(\ref{eq:terms}), and they assumed that the crest
extended as $\rho_\chi\propto r^{-3/2}$ all the way to $r_h$,
rather than to only a fraction of $r_h$.

The relative contribution from the crest could be increased
by assuming a steeper DM density falloff,
e.g. in a halo
where the DM had been pulled in by contraction of the baryons,
or a smaller angular acceptance $\Delta\Omega$ of the detector.
As discussed above, observations favor a {\it lower}
DM density implying even less of a contribution from 
the crest.

\begin{figure}
\includegraphics[width=0.425\textwidth]{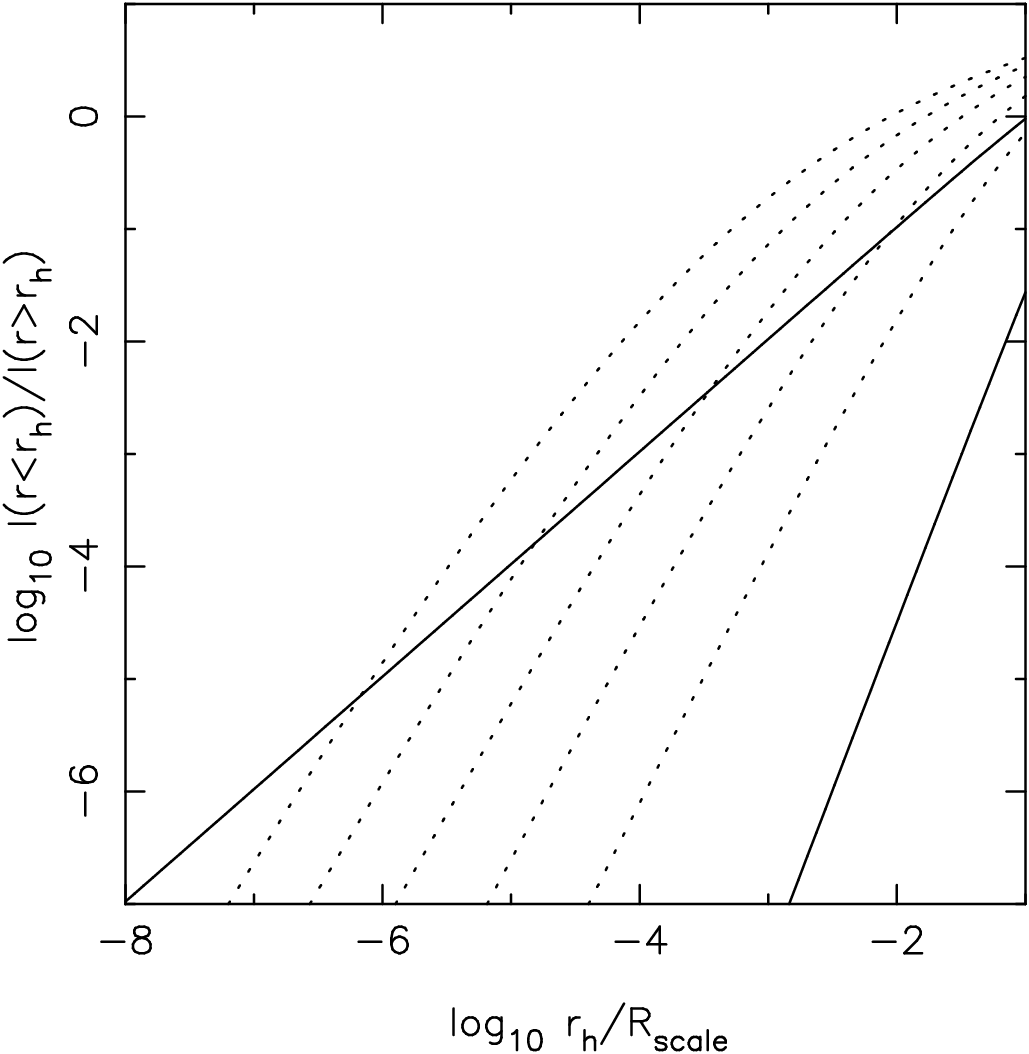}
\caption{\label{fig:ext} 
Contribution to the self-annihilation signal from DM
at $r\le r_h$, i.e. within the SBH's influence radius.
The DM distribution at $r\le r_h$ is assumed to be the
same as in Fig.~\ref{fig:NB} at $t=T_r(r_h)$.
Three different models for the DM density at
$r>r_h$ have been evaluated:
{\it Lower solid line:} Burkert profile;
{\it Upper solid line:} NFW profile;
{\it Dotted lines:} Prugniel-Simien profile, with
$n=(2,3,4,5,6)$, increasing upwards.
$R_{scale}$ is the appropriate scale length for
each model, i.e. $r_c$ (Burkert), $r_s$ (NFW),
$R_e$ (Prugniel-Simien).
}
\end{figure}

\subsection{External Galaxies}

In the case of a detector centered on the SBH of an external galaxy,
the geometrical term in the expression for the flux becomes
\begin{equation}
I = {1\over D^2} \int \rho_\chi^2(r) r^2 dr 
\label{eq:fluxext}
\end{equation}
where $D$ is the distance to the galaxy and the integral
includes as much of the galaxy as is imaged by the telescope;
thus
\begin{equation}
I(\Psi) = {1\over D^2}
\left [ \int_{r_0}^{r_h} \rho_\chi^2(r)r^2dr 
+  \int_{r_h}^{\Psi D} \rho_\chi^2(r) r^2 dr \right ] .
\label{eq:termsext}
\end{equation}
We again equate the first term in brackets with
Eq.~(\ref{eq:intb}).
For the low-luminosity galaxies that are likely to harbor
a DM
crest (Sect. IVA),
the gamma-ray telescope would image essentially the entire halo
and so we set the upper limit of the second integral
to infinity in what follows.

Following the discussion in Sect. IIIC,
we considered three possible forms for the DM density
beyond $r= r_h$:
the NFW profile (Eq.~\ref{eq:NFW}),
the Burkert profile (Eq.~\ref{eq:Burkert})
and the Prugniel-Simien profile (Eq.~\ref{eq:PS}).

Fig.~\ref{fig:ext} shows the relative contribution
of  the DM at $r\le r_h$ to the annihilation signal
for the three assumed DM profiles.
In the case of the Prugniel-Simien profile,
the curvature parameter $n$ has been varied over
the range $2\le n\le 6$ that approximately characterizes
the $N$-body haloes \cite{Empirical1}.
The contribution of  the crest is insignificant
in the  case of the Burkert profile and only
significant in the  case of the NFW profile
if $r_h/r_s$ is unphysically large.
For the Prugniel-Simien profile however,
the crest adds significantly to the total signal
for large $n$ and $r_h/R_e\gtrsim 10^{-4}$.

Whether such large values of $r_h/R_e$ are physically
reasonable depends on the poorly-understood relations
between SBH mass, galaxy mass and DM halo properties
at the low-mass end of these distributions.
The scale radii $R_e$ of $N$-body DM halos obey
\begin{equation}
{R_e\over 100\ {\rm kpc}} \approx 1.1 \left({M_{DM}\over 10^{12}M_\odot}\right)^{1/2}
\end{equation}
\cite{Empirical3}.
Thus
\begin{eqnarray}
{r_h\over R_e} &\approx&{GM_\bullet\over\sigma^2_\star R_e} \nonumber \\
&\approx& 4\times 10^{-6} \left({M_\bullet\over 10^6M_\odot}\right)
\left({\sigma_\star\over 100\ {\rm km\ s}^{-1}}\right)^{-2}
\left({M_{DM}\over 10^{12}M_\odot}\right)^{-1/2}.
\end{eqnarray}
Adopting the empirical relation between $M_\bullet$ and $\sigma_\star$
\cite{FF-05} 
\begin{equation}
{M_\bullet\over 10^6M_\odot} \approx 5.7\left({\sigma_\star\over 100\ {\rm km\ s}^{-1}}\right)^\alpha,\ \ \ \ \alpha\approx 4.86
\end{equation}
(which has only been established for
$\sigma_\star\gtrsim 100\ {\rm km\ s}^{-1}$)
this becomes
\begin{equation}
{r_h\over R_e} \approx 8\times 10^{-6} 
\left({M_\bullet\over 10^6M_\odot}\right)^{0.59}
\left({M_{DM}\over 10^{12}M_\odot}\right)^{-1/2}.
\end{equation}
If $M_\bullet$ scales linearly with halo mass
at the low-mass end of the distribution,
this relation implies that $r_h/R_e$ is essentially
independent of $M_\bullet$ and $M_{DM}$, hence
roughly equal to its value in the Milky Way,
$r_h/R_e\approx 1\times 10^{-5}$.

On the other hand, 
the empirical relation between rotation curve
peak velocity and stellar velocity dispersion
suggests a nonlinear relation between $M_\bullet$
and $M_{DM}$, of the approximate form \cite{Ferrarese-02}
\begin{equation}
{M_\bullet\over 10^6M_\odot} \approx 
K\left({M_{DM}\over 10^{12}M_\odot}\right)^\beta
\end{equation}
with $1.5\lesssim\beta\lesssim 2$
and $0.1\lesssim K \lesssim 1$.
Thus
\begin{equation}
{r_h\over R_e} \propto M_\bullet^{0.59-1/2\beta}
\propto M_{DM}^{0.59\beta-1/2}
\end{equation}
and $r_h/R_e$ would scale approximately as
$M_\bullet^{0.3}\propto M_{DM}^{0.6}$.

These scaling relations should be considered
highly uncertain for the reasons discussed in Sect. IV.
A third possibility, not inconsistent with the limited
observational constraints on SBH and halo masses,
is that SBH mass is essentially 
independent of halo mass at the low-mass end of the
distributions.
This assumption would imply increasing values
of $r_h/R_e$ in dwarf galaxies, hence larger
contributions from the crests to the annihilation
signal (Fig.~\ref{fig:ext}).

\subsection{Dependence on initial conditions and galaxy age}

The estimates made above of DM crest observability were based
on the $N$-body density profile, Fig.~\ref{fig:NB}, at
$t=T_r(r_h)$.
While this density profile is ``universal'' at $r\lesssim 0.1 r_h$
in the sense that $\rho_\chi\sim r^{-3/2}$ is a steady-state
solution to the Fokker-Planck equation in a point-mass potential,
the amplitude of the crest decays after its formation due
to continued heating by the stars (Fig.~\ref{fig:fp2}; 
\cite{Merritt-04,BM-05}).
Furthermore 
the distribution of DM within the SBH's influence radius
could be different if the initial conditions were very
different from those assumed here.

{\it Initial conditions:} Until now we have made the conservative
assumption that the DM density profile was initially very flat
near the center.
These are appropriate initial conditions if the galaxy hosting
the DM experienced a merger following formation of
SBHs, since a binary SBH is efficient at ejecting matter from
the center of a galaxy and creating a low-density core
\cite{MMVJ-02}.
However the mean time between mergers is a strong function
of galaxy luminosity/mass, and an isolated, low-mass stellar
spheroid might not have experienced a significant merger
in the last $10^{10}$ yr. 
(We note that this statement does not apply to the bulges
of massive spiral galaxies, like that of the Milky Way, since the
merger probability is determined by the overall mass/radius
of the galaxy, and these are much larger than the mass/radius
of the bulge in the case of a spiral galaxy.)
For instance, intermediate-mass black holes (IMBHs) might have
formed via direct collapse of primoridial gas in low-mass
halos \cite{KBD-04}.
In this scenario, the distribution of DM around the IMBH
could have the steep dependence with radius predicted by
so-called ``adiabatic growth'' models,
$\rho_\chi\sim r^{-\gamma}, 2\lesssim\gamma\lesssim 3$ 
\cite{Bertone:2005xz,Bertone:2006nq}.
Such steep spikes would be ``softened'' by self-annihilations
at small radii and by heating from the stars, but Fokker-Planck
integrations show that the density profile after one relaxation
time can remain considerably steeper than that of the collisional,
$\rho_\chi\sim r^{-3/2}$ crests discussed here \cite{Merritt-04,BM-05}.
The same considerations would apply to a spike in a more
massive halo that did not happen to experience a major merger
since the epoch of SBH formation.

{\it Age:} On time scales long compared with $T_r(r_h)$,
continued collisional evolution of a nucleus can 
result in different stellar and/or DM density profiles at 
$r\lesssim r_h$.
Heating of DM by stars causes the
normalization of the DM density to drop, while roughly
maintaining the $\rho_\chi\sim r^{-3/2}$ dependence
at $r\ll r_h$ (e.g. Fig.~ 1, 2).
%Other effects associated with star-star interactions
%can drive evolution of a stellar nucleus on time
%scales long compared with $T_r(r_h)$, indirectly
%affecting the DM distribution.
Capture or tidal disruption of stars by the SBH is
effectively a heat source, causing a nucleus to 
expand \cite{Shapiro-77}; the time scale is of
order $\sim T_r(r_h)$ and depends on the sizes and
masses of stars.
If there is a range of stellar masses,
the more massive stars will
accumulate near the center, causing the stellar mass
density profile to become more centrally peaked \cite{RPP-06}.
While the ``steady-state'' functional form for $\rho_\chi(r)$,
Eq.~(\ref{eq:ganal}),
is formally independent of $\rho_\star(r)$, the rate at which
the stars transfer energy to the DM depends on the
stellar density.
As discussed above (Sect. IV A), it is not clear that any
nucleus harboring a SBH is much older than one relaxation time.

\section{\label{sec:sum}Summary}

By considering the joint evolution of the stellar and dark-matter
(DM) densities at the center of a galaxy containing a supermassive
black hole (SBH), we have shown that the existence of a ``crest'' 
(collisionally regenerated dark-matter structure)
can be robustly predicted in any nucleus old enough to have
generated a Bahcall-Wolf cusp in the stars.
This time scale is roughly $10$ Gyr in the case of the Milky
Way, and probably shorter for fainter galaxies that contain massive BHs.
Crest generation occurs even in the (probably generic) case of a nucleus 
that previously experienced the scouring effect of a binary
SBH.
Standard galaxy scaling relations suggest that crests do not 
dramatically change the prospects for indirect detection of DM.
However this conclusion could be modified if the
DM density falls steeply beyond the SBH's 
gravitational influence radius, or if SBH masses scale weakly
with DM halo masses.
In any case, crest formation implies a significant increase
in the density of DM around the central SBH, and
the presence of crests should be taken into account when 
studying the constraints on the annihilation rate from stellar orbits
\cite{Hall:2006na}, or the evolution of stars in the innermost 
regions of galactic nuclei \cite{Salati89,Moskalenko:2006mk}.

\bigskip\bigskip

\section{Acknowledgements}

We thank the organizers of the KITP workshop on
``Physics of Galactic Nuclei'' where this work was begun. 
This research was supported in part 
by the National Science Foundation under grants no. PHY99-07949,
AST-0206031, AST-0420920 and AST-0437519, by the National
Aeronautics and Space Administration under grant no. NNG04GJ48G,
and by the Space Telescope Science Institute under 
grant no. HST-AR-09519.01-A.
GB is supported by the Helmholtz Association of National Research 
Centres, under project VH-NG-006. 
The $N$-body calculations presented here were carried out at
the Center for the Advancement of the Study of 
Cyberinfrastructure at RIT whose support is gratefully
acknowledged.

\bibliography{ms}

\begin{thebibliography}{92}
\expandafter\ifx\csname natexlab\endcsname\relax\def\natexlab#1{#1}\fi
\expandafter\ifx\csname bibnamefont\endcsname\relax
  \def\bibnamefont#1{#1}\fi
\expandafter\ifx\csname bibfnamefont\endcsname\relax
  \def\bibfnamefont#1{#1}\fi
\expandafter\ifx\csname citenamefont\endcsname\relax
  \def\citenamefont#1{#1}\fi
\expandafter\ifx\csname url\endcsname\relax
  \def\url#1{\texttt{#1}}\fi
\expandafter\ifx\csname urlprefix\endcsname\relax\def\urlprefix{URL }\fi
\providecommand{\bibinfo}[2]{#2}
\providecommand{\eprint}[2][]{\url{#2}}

\bibitem[{\citenamefont{{Bertone} et~al.}(2004)\citenamefont{{Bertone},
  {Hooper}, and {Silk}}}]{BHS-04}
\bibinfo{author}{\bibfnamefont{G.}~\bibnamefont{{Bertone}}},
  \bibinfo{author}{\bibfnamefont{D.}~\bibnamefont{{Hooper}}}, \bibnamefont{and}
  \bibinfo{author}{\bibfnamefont{J.}~\bibnamefont{{Silk}}},
  \bibinfo{journal}{Physics Reports} \textbf{\bibinfo{volume}{405}},
  \bibinfo{pages}{279} (\bibinfo{year}{2004}).

\bibitem[{\citenamefont{Bergstrom}(2000)}]{Bergstrom:2000pn}
\bibinfo{author}{\bibfnamefont{L.}~\bibnamefont{Bergstrom}},
  \bibinfo{journal}{Rept. Prog. Phys.} \textbf{\bibinfo{volume}{63}},
  \bibinfo{pages}{793} (\bibinfo{year}{2000}), \eprint{hep-ph/0002126}.

\bibitem[{\citenamefont{{Bertone} and {Merritt}}(2005{\natexlab{a}})}]{MPLA}
\bibinfo{author}{\bibfnamefont{G.}~\bibnamefont{{Bertone}}} \bibnamefont{and}
  \bibinfo{author}{\bibfnamefont{D.}~\bibnamefont{{Merritt}}},
  \bibinfo{journal}{Modern Physics Letters A} \textbf{\bibinfo{volume}{20}},
  \bibinfo{pages}{1021} (\bibinfo{year}{2005}{\natexlab{a}}).

\bibitem[{\citenamefont{{Carr} et~al.}(2006)\citenamefont{{Carr}, {Lamanna},
  and {Lavalle}}}]{Carr-06}
\bibinfo{author}{\bibfnamefont{J.}~\bibnamefont{{Carr}}},
  \bibinfo{author}{\bibfnamefont{G.}~\bibnamefont{{Lamanna}}},
  \bibnamefont{and}
  \bibinfo{author}{\bibfnamefont{J.}~\bibnamefont{{Lavalle}}},
  \bibinfo{journal}{Reports of Progress in Physics}
  \textbf{\bibinfo{volume}{69}}, \bibinfo{pages}{2475} (\bibinfo{year}{2006}).

\bibitem[{\citenamefont{{Stecker}}(1988)}]{Stecker-88}
\bibinfo{author}{\bibfnamefont{F.~W.} \bibnamefont{{Stecker}}},
  \bibinfo{journal}{Physics Letters B} \textbf{\bibinfo{volume}{201}},
  \bibinfo{pages}{529} (\bibinfo{year}{1988}).

\bibitem[{\citenamefont{{Bergstr{\"o}m}
  et~al.}(1998)\citenamefont{{Bergstr{\"o}m}, {Ullio}, and {Buckley}}}]{BUB-98}
\bibinfo{author}{\bibfnamefont{L.}~\bibnamefont{{Bergstr{\"o}m}}},
  \bibinfo{author}{\bibfnamefont{P.}~\bibnamefont{{Ullio}}}, \bibnamefont{and}
  \bibinfo{author}{\bibfnamefont{J.~H.} \bibnamefont{{Buckley}}},
  \bibinfo{journal}{Astroparticle Physics} \textbf{\bibinfo{volume}{9}},
  \bibinfo{pages}{137} (\bibinfo{year}{1998}).

\bibitem[{\citenamefont{{Bertone} et~al.}(2001)\citenamefont{{Bertone}, {Sigl},
  and {Silk}}}]{BSS-01}
\bibinfo{author}{\bibfnamefont{G.}~\bibnamefont{{Bertone}}},
  \bibinfo{author}{\bibfnamefont{G.}~\bibnamefont{{Sigl}}}, \bibnamefont{and}
  \bibinfo{author}{\bibfnamefont{J.}~\bibnamefont{{Silk}}},
  \bibinfo{journal}{Monthly Notices of the Royal Astronomical Society}
  \textbf{\bibinfo{volume}{326}}, \bibinfo{pages}{799} (\bibinfo{year}{2001}).

\bibitem[{\citenamefont{{Fornengo} et~al.}(2004)\citenamefont{{Fornengo},
  {Pieri}, and {Scopel}}}]{FPS-04}
\bibinfo{author}{\bibfnamefont{N.}~\bibnamefont{{Fornengo}}},
  \bibinfo{author}{\bibfnamefont{L.}~\bibnamefont{{Pieri}}}, \bibnamefont{and}
  \bibinfo{author}{\bibfnamefont{S.}~\bibnamefont{{Scopel}}},
  \bibinfo{journal}{\prd} \textbf{\bibinfo{volume}{70}},
  \bibinfo{pages}{103529} (\bibinfo{year}{2004}).

\bibitem[{\citenamefont{{Hooper} et~al.}(2004)\citenamefont{{Hooper}, {de la
  Calle Perez}, {Silk}, {Ferrer}, and {Sarkar}}}]{Hooper-04}
\bibinfo{author}{\bibfnamefont{D.}~\bibnamefont{{Hooper}}},
  \bibinfo{author}{\bibfnamefont{I.}~\bibnamefont{{de la Calle Perez}}},
  \bibinfo{author}{\bibfnamefont{J.}~\bibnamefont{{Silk}}},
  \bibinfo{author}{\bibfnamefont{F.}~\bibnamefont{{Ferrer}}}, \bibnamefont{and}
  \bibinfo{author}{\bibfnamefont{S.}~\bibnamefont{{Sarkar}}},
  \bibinfo{journal}{Journal of Cosmology and Astro-Particle Physics}
  \textbf{\bibinfo{volume}{9}}, \bibinfo{pages}{2} (\bibinfo{year}{2004}).

\bibitem[{\citenamefont{{Ferrarese} and {Ford}}(2005)}]{FF-05}
\bibinfo{author}{\bibfnamefont{L.}~\bibnamefont{{Ferrarese}}} \bibnamefont{and}
  \bibinfo{author}{\bibfnamefont{H.}~\bibnamefont{{Ford}}},
  \bibinfo{journal}{Space Science Reviews} \textbf{\bibinfo{volume}{116}},
  \bibinfo{pages}{523} (\bibinfo{year}{2005}).

\bibitem[{\citenamefont{{Merritt}}(2006)}]{RPP-06}
\bibinfo{author}{\bibfnamefont{D.}~\bibnamefont{{Merritt}}},
  \bibinfo{journal}{Reports on Progress in Physics}
  \textbf{\bibinfo{volume}{69}}, \bibinfo{pages}{2513} (\bibinfo{year}{2006}).

\bibitem[{\citenamefont{{Gondolo} and {Silk}}(1999)}]{GS-99}
\bibinfo{author}{\bibfnamefont{P.}~\bibnamefont{{Gondolo}}} \bibnamefont{and}
  \bibinfo{author}{\bibfnamefont{J.}~\bibnamefont{{Silk}}},
  \bibinfo{journal}{Physical Review Letters} \textbf{\bibinfo{volume}{83}},
  \bibinfo{pages}{1719} (\bibinfo{year}{1999}).

\bibitem[{\citenamefont{Bertone et~al.}(2002)\citenamefont{Bertone, Sigl, and
  Silk}}]{Bertone:2002je}
\bibinfo{author}{\bibfnamefont{G.}~\bibnamefont{Bertone}},
  \bibinfo{author}{\bibfnamefont{G.}~\bibnamefont{Sigl}}, \bibnamefont{and}
  \bibinfo{author}{\bibfnamefont{J.}~\bibnamefont{Silk}},
  \bibinfo{journal}{Mon. Not. Roy. Astron. Soc.}
  \textbf{\bibinfo{volume}{337}}, \bibinfo{pages}{98} (\bibinfo{year}{2002}),
  \eprint{astro-ph/0203488}.

\bibitem[{\citenamefont{{Aloisio} et~al.}(2004)\citenamefont{{Aloisio},
  {Blasi}, and {Olinto}}}]{ABO-04}
\bibinfo{author}{\bibfnamefont{R.}~\bibnamefont{{Aloisio}}},
  \bibinfo{author}{\bibfnamefont{P.}~\bibnamefont{{Blasi}}}, \bibnamefont{and}
  \bibinfo{author}{\bibfnamefont{A.~V.} \bibnamefont{{Olinto}}},
  \bibinfo{journal}{Journal of Cosmology and Astro-Particle Physics}
  \textbf{\bibinfo{volume}{5}}, \bibinfo{pages}{7} (\bibinfo{year}{2004}).

\bibitem[{\citenamefont{{Ullio} et~al.}(2001)\citenamefont{{Ullio}, {Zhao}, and
  {Kamionkowski}}}]{UZK-01}
\bibinfo{author}{\bibfnamefont{P.}~\bibnamefont{{Ullio}}},
  \bibinfo{author}{\bibfnamefont{H.}~\bibnamefont{{Zhao}}}, \bibnamefont{and}
  \bibinfo{author}{\bibfnamefont{M.}~\bibnamefont{{Kamionkowski}}},
  \bibinfo{journal}{\prd} \textbf{\bibinfo{volume}{64}},
  \bibinfo{pages}{043504} (\bibinfo{year}{2001}).

\bibitem[{\citenamefont{{Merritt} et~al.}(2002)\citenamefont{{Merritt},
  {Milosavljevi{\'c}}, {Verde}, and {Jimenez}}}]{MMVJ-02}
\bibinfo{author}{\bibfnamefont{D.}~\bibnamefont{{Merritt}}},
  \bibinfo{author}{\bibfnamefont{M.}~\bibnamefont{{Milosavljevi{\'c}}}},
  \bibinfo{author}{\bibfnamefont{L.}~\bibnamefont{{Verde}}}, \bibnamefont{and}
  \bibinfo{author}{\bibfnamefont{R.}~\bibnamefont{{Jimenez}}},
  \bibinfo{journal}{Physical Review Letters} \textbf{\bibinfo{volume}{88}},
  \bibinfo{pages}{191301} (\bibinfo{year}{2002}).

\bibitem[{\citenamefont{{Bertone} and {Merritt}}(2005{\natexlab{b}})}]{BM-05}
\bibinfo{author}{\bibfnamefont{G.}~\bibnamefont{{Bertone}}} \bibnamefont{and}
  \bibinfo{author}{\bibfnamefont{D.}~\bibnamefont{{Merritt}}},
  \bibinfo{journal}{\prd} \textbf{\bibinfo{volume}{72}},
  \bibinfo{pages}{103502} (\bibinfo{year}{2005}{\natexlab{b}}).

\bibitem[{\citenamefont{{Merritt}}(2005)}]{Merritt-06}
\bibinfo{author}{\bibfnamefont{D.}~\bibnamefont{{Merritt}}},
  \bibinfo{journal}{The Astrophysical Journal} \textbf{\bibinfo{volume}{648}},
  \bibinfo{pages}{000} (\bibinfo{year}{2005}).

\bibitem[{\citenamefont{{Lauer} et~al.}(1998)\citenamefont{{Lauer}, {Faber},
  {Ajhar}, {Grillmair}, and {Scowen}}}]{Lauer-98}
\bibinfo{author}{\bibfnamefont{T.~R.} \bibnamefont{{Lauer}}},
  \bibinfo{author}{\bibfnamefont{S.~M.} \bibnamefont{{Faber}}},
  \bibinfo{author}{\bibfnamefont{E.~A.} \bibnamefont{{Ajhar}}},
  \bibinfo{author}{\bibfnamefont{C.~J.} \bibnamefont{{Grillmair}}},
  \bibnamefont{and} \bibinfo{author}{\bibfnamefont{P.~A.}
  \bibnamefont{{Scowen}}}, \bibinfo{journal}{Astronomical Journal}
  \textbf{\bibinfo{volume}{116}}, \bibinfo{pages}{2263} (\bibinfo{year}{1998}).

\bibitem[{\citenamefont{{Alexander}}(1999)}]{Alexander-99}
\bibinfo{author}{\bibfnamefont{T.}~\bibnamefont{{Alexander}}},
  \bibinfo{journal}{\apj} \textbf{\bibinfo{volume}{527}}, \bibinfo{pages}{835}
  (\bibinfo{year}{1999}).

\bibitem[{\citenamefont{{Bahcall} and {Wolf}}(1976)}]{BW-76}
\bibinfo{author}{\bibfnamefont{J.~N.} \bibnamefont{{Bahcall}}}
  \bibnamefont{and} \bibinfo{author}{\bibfnamefont{R.~A.}
  \bibnamefont{{Wolf}}}, \bibinfo{journal}{\apj}
  \textbf{\bibinfo{volume}{209}}, \bibinfo{pages}{214} (\bibinfo{year}{1976}).

\bibitem[{\citenamefont{{Merritt} and {Szell}}(2005)}]{MS-06}
\bibinfo{author}{\bibfnamefont{D.}~\bibnamefont{{Merritt}}} \bibnamefont{and}
  \bibinfo{author}{\bibfnamefont{A.}~\bibnamefont{{Szell}}},
  \bibinfo{journal}{The Astrophysical Journal} \textbf{\bibinfo{volume}{648}},
  \bibinfo{pages}{000} (\bibinfo{year}{2005}).

\bibitem[{\citenamefont{{Ilyin} et~al.}(2004)\citenamefont{{Ilyin}, {Zybin},
  and {Gurevich}}}]{IZG-04}
\bibinfo{author}{\bibfnamefont{A.~S.} \bibnamefont{{Ilyin}}},
  \bibinfo{author}{\bibfnamefont{K.~P.} \bibnamefont{{Zybin}}},
  \bibnamefont{and} \bibinfo{author}{\bibfnamefont{A.~V.}
  \bibnamefont{{Gurevich}}}, \bibinfo{journal}{Journal of Experimental and
  Theoretical Physics} \textbf{\bibinfo{volume}{98}}, \bibinfo{pages}{1}
  (\bibinfo{year}{2004}).

\bibitem[{\citenamefont{{Merritt}}(2004)}]{Merritt-04}
\bibinfo{author}{\bibfnamefont{D.}~\bibnamefont{{Merritt}}},
  \bibinfo{journal}{Physical Review Letters} \textbf{\bibinfo{volume}{92}},
  \bibinfo{pages}{201304} (\bibinfo{year}{2004}).

\bibitem[{\citenamefont{{Gnedin} and {Primack}}(2004)}]{GP-04}
\bibinfo{author}{\bibfnamefont{O.~Y.} \bibnamefont{{Gnedin}}} \bibnamefont{and}
  \bibinfo{author}{\bibfnamefont{J.~R.} \bibnamefont{{Primack}}},
  \bibinfo{journal}{Physical Review Letters} \textbf{\bibinfo{volume}{93}},
  \bibinfo{pages}{061302} (\bibinfo{year}{2004}).

\bibitem[{\citenamefont{{H{\'e}non}}(1961)}]{Henon-61}
\bibinfo{author}{\bibfnamefont{M.}~\bibnamefont{{H{\'e}non}}},
  \bibinfo{journal}{Annales d'Astrophysique} \textbf{\bibinfo{volume}{24}},
  \bibinfo{pages}{369} (\bibinfo{year}{1961}).

\bibitem[{\citenamefont{{Spitzer}}(1987)}]{Spitzer-87}
\bibinfo{author}{\bibfnamefont{L.}~\bibnamefont{{Spitzer}}},
  \emph{\bibinfo{title}{{Dynamical evolution of globular clusters}}}
  (\bibinfo{publisher}{Princeton, NJ, Princeton University Press, 191 p.},
  \bibinfo{year}{1987}).

\bibitem[{\citenamefont{{Merritt}}(1983)}]{Merritt-83}
\bibinfo{author}{\bibfnamefont{D.}~\bibnamefont{{Merritt}}},
  \bibinfo{journal}{\apj} \textbf{\bibinfo{volume}{264}}, \bibinfo{pages}{24}
  (\bibinfo{year}{1983}).

\bibitem[{\citenamefont{{Bahcall} and {Wolf}}(1977)}]{BW-77}
\bibinfo{author}{\bibfnamefont{J.~N.} \bibnamefont{{Bahcall}}}
  \bibnamefont{and} \bibinfo{author}{\bibfnamefont{R.~A.}
  \bibnamefont{{Wolf}}}, \bibinfo{journal}{\apj}
  \textbf{\bibinfo{volume}{216}}, \bibinfo{pages}{883} (\bibinfo{year}{1977}).

\bibitem[{\citenamefont{{Dehnen}}(1993)}]{Dehnen-93}
\bibinfo{author}{\bibfnamefont{W.}~\bibnamefont{{Dehnen}}},
  \bibinfo{journal}{Monthly Notices of the Royal Astronomical Society}
  \textbf{\bibinfo{volume}{265}}, \bibinfo{pages}{250} (\bibinfo{year}{1993}).

\bibitem[{\citenamefont{{Harfst} et~al.}(2006)\citenamefont{{Harfst},
  {Gualandris}, {Merritt}, {Spurzem}, {Portegies Zwart}, and
  {Berczik}}}]{Harfst-06}
\bibinfo{author}{\bibfnamefont{S.}~\bibnamefont{{Harfst}}},
  \bibinfo{author}{\bibfnamefont{A.}~\bibnamefont{{Gualandris}}},
  \bibinfo{author}{\bibfnamefont{D.}~\bibnamefont{{Merritt}}},
  \bibinfo{author}{\bibfnamefont{R.}~\bibnamefont{{Spurzem}}},
  \bibinfo{author}{\bibfnamefont{S.}~\bibnamefont{{Portegies Zwart}}},
  \bibnamefont{and}
  \bibinfo{author}{\bibfnamefont{P.}~\bibnamefont{{Berczik}}},
  \bibinfo{journal}{ArXiv Astrophysics e-prints}  (\bibinfo{year}{2006}).

\bibitem[{\citenamefont{{Makino} and {Taiji}}(1998)}]{GRAPE}
\bibinfo{author}{\bibfnamefont{J.}~\bibnamefont{{Makino}}} \bibnamefont{and}
  \bibinfo{author}{\bibfnamefont{M.}~\bibnamefont{{Taiji}}},
  \emph{\bibinfo{title}{{Scientific simulations with special-purpose computers
  : The GRAPE systems}}} (\bibinfo{publisher}{Toronto, John Wiley and Sons},
  \bibinfo{year}{1998}).

\bibitem[{\citenamefont{{Ferrarese}
  et~al.}(2006{\natexlab{a}})\citenamefont{{Ferrarese}, {C{\^o}t{\'e}},
  {Jord{\'a}n}, {Peng}, {Blakeslee}, {Piatek}, {Mei}, {Merritt},
  {Milosavljevi{\'c}}, {Tonry} et~al.}}]{ACS6}
\bibinfo{author}{\bibfnamefont{L.}~\bibnamefont{{Ferrarese}}},
  \bibinfo{author}{\bibfnamefont{P.}~\bibnamefont{{C{\^o}t{\'e}}}},
  \bibinfo{author}{\bibfnamefont{A.}~\bibnamefont{{Jord{\'a}n}}},
  \bibinfo{author}{\bibfnamefont{E.~W.} \bibnamefont{{Peng}}},
  \bibinfo{author}{\bibfnamefont{J.~P.} \bibnamefont{{Blakeslee}}},
  \bibinfo{author}{\bibfnamefont{S.}~\bibnamefont{{Piatek}}},
  \bibinfo{author}{\bibfnamefont{S.}~\bibnamefont{{Mei}}},
  \bibinfo{author}{\bibfnamefont{D.}~\bibnamefont{{Merritt}}},
  \bibinfo{author}{\bibfnamefont{M.}~\bibnamefont{{Milosavljevi{\'c}}}},
  \bibinfo{author}{\bibfnamefont{J.~L.} \bibnamefont{{Tonry}}},
  \bibnamefont{et~al.}, \bibinfo{journal}{\apj Supplement}
  \textbf{\bibinfo{volume}{164}}, \bibinfo{pages}{334}
  (\bibinfo{year}{2006}{\natexlab{a}}).

\bibitem[{\citenamefont{{Ghez} et~al.}(2005)\citenamefont{{Ghez}, {Salim},
  {Hornstein}, {Tanner}, {Lu}, {Morris}, {Becklin}, and
  {Duch{\^e}ne}}}]{Ghez-05}
\bibinfo{author}{\bibfnamefont{A.~M.} \bibnamefont{{Ghez}}},
  \bibinfo{author}{\bibfnamefont{S.}~\bibnamefont{{Salim}}},
  \bibinfo{author}{\bibfnamefont{S.~D.} \bibnamefont{{Hornstein}}},
  \bibinfo{author}{\bibfnamefont{A.}~\bibnamefont{{Tanner}}},
  \bibinfo{author}{\bibfnamefont{J.~R.} \bibnamefont{{Lu}}},
  \bibinfo{author}{\bibfnamefont{M.}~\bibnamefont{{Morris}}},
  \bibinfo{author}{\bibfnamefont{E.~E.} \bibnamefont{{Becklin}}},
  \bibnamefont{and}
  \bibinfo{author}{\bibfnamefont{G.}~\bibnamefont{{Duch{\^e}ne}}},
  \bibinfo{journal}{\apj} \textbf{\bibinfo{volume}{620}}, \bibinfo{pages}{744}
  (\bibinfo{year}{2005}).

\bibitem[{\citenamefont{{Preto} et~al.}(2004)\citenamefont{{Preto}, {Merritt},
  and {Spurzem}}}]{Preto-04}
\bibinfo{author}{\bibfnamefont{M.}~\bibnamefont{{Preto}}},
  \bibinfo{author}{\bibfnamefont{D.}~\bibnamefont{{Merritt}}},
  \bibnamefont{and}
  \bibinfo{author}{\bibfnamefont{R.}~\bibnamefont{{Spurzem}}},
  \bibinfo{journal}{\apj Letters} \textbf{\bibinfo{volume}{613}},
  \bibinfo{pages}{L109} (\bibinfo{year}{2004}).

\bibitem[{\citenamefont{{Baumgardt}
  et~al.}(2004{\natexlab{a}})\citenamefont{{Baumgardt}, {Makino}, and
  {Ebisuzaki}}}]{Baumgardt-04}
\bibinfo{author}{\bibfnamefont{H.}~\bibnamefont{{Baumgardt}}},
  \bibinfo{author}{\bibfnamefont{J.}~\bibnamefont{{Makino}}}, \bibnamefont{and}
  \bibinfo{author}{\bibfnamefont{T.}~\bibnamefont{{Ebisuzaki}}},
  \bibinfo{journal}{\apj} \textbf{\bibinfo{volume}{613}}, \bibinfo{pages}{1133}
  (\bibinfo{year}{2004}{\natexlab{a}}).

\bibitem[{\citenamefont{{C{\^o}t{\'e}}
  et~al.}(2006)\citenamefont{{C{\^o}t{\'e}}, {Piatek}, {Ferrarese},
  {Jord{\'a}n}, {Merritt}, {Peng}, {Ha{\c s}egan}, {Blakeslee}, {Mei}, {West}
  et~al.}}]{ACS8}
\bibinfo{author}{\bibfnamefont{P.}~\bibnamefont{{C{\^o}t{\'e}}}},
  \bibinfo{author}{\bibfnamefont{S.}~\bibnamefont{{Piatek}}},
  \bibinfo{author}{\bibfnamefont{L.}~\bibnamefont{{Ferrarese}}},
  \bibinfo{author}{\bibfnamefont{A.}~\bibnamefont{{Jord{\'a}n}}},
  \bibinfo{author}{\bibfnamefont{D.}~\bibnamefont{{Merritt}}},
  \bibinfo{author}{\bibfnamefont{E.~W.} \bibnamefont{{Peng}}},
  \bibinfo{author}{\bibfnamefont{M.}~\bibnamefont{{Ha{\c s}egan}}},
  \bibinfo{author}{\bibfnamefont{J.~P.} \bibnamefont{{Blakeslee}}},
  \bibinfo{author}{\bibfnamefont{S.}~\bibnamefont{{Mei}}},
  \bibinfo{author}{\bibfnamefont{M.~J.} \bibnamefont{{West}}},
  \bibnamefont{et~al.}, \bibinfo{journal}{\apj Supplement}
  \textbf{\bibinfo{volume}{165}}, \bibinfo{pages}{57} (\bibinfo{year}{2006}).

\bibitem[{\citenamefont{{Baumgardt}
  et~al.}(2004{\natexlab{b}})\citenamefont{{Baumgardt}, {Makino}, and
  {Ebisuzaki}}}]{Baumgardt-04b}
\bibinfo{author}{\bibfnamefont{H.}~\bibnamefont{{Baumgardt}}},
  \bibinfo{author}{\bibfnamefont{J.}~\bibnamefont{{Makino}}}, \bibnamefont{and}
  \bibinfo{author}{\bibfnamefont{T.}~\bibnamefont{{Ebisuzaki}}},
  \bibinfo{journal}{\apj} \textbf{\bibinfo{volume}{613}}, \bibinfo{pages}{1143}
  (\bibinfo{year}{2004}{\natexlab{b}}).

\bibitem[{\citenamefont{{B{\"o}ker} et~al.}(2004)\citenamefont{{B{\"o}ker},
  {Sarzi}, {McLaughlin}, {van der Marel}, {Rix}, {Ho}, and
  {Shields}}}]{Boeker-04}
\bibinfo{author}{\bibfnamefont{T.}~\bibnamefont{{B{\"o}ker}}},
  \bibinfo{author}{\bibfnamefont{M.}~\bibnamefont{{Sarzi}}},
  \bibinfo{author}{\bibfnamefont{D.~E.} \bibnamefont{{McLaughlin}}},
  \bibinfo{author}{\bibfnamefont{R.~P.} \bibnamefont{{van der Marel}}},
  \bibinfo{author}{\bibfnamefont{H.-W.} \bibnamefont{{Rix}}},
  \bibinfo{author}{\bibfnamefont{L.~C.} \bibnamefont{{Ho}}}, \bibnamefont{and}
  \bibinfo{author}{\bibfnamefont{J.~C.} \bibnamefont{{Shields}}},
  \bibinfo{journal}{Astronomical Journal} \textbf{\bibinfo{volume}{127}},
  \bibinfo{pages}{105} (\bibinfo{year}{2004}).

\bibitem[{\citenamefont{{Valluri} et~al.}(2005)\citenamefont{{Valluri},
  {Ferrarese}, {Merritt}, and {Joseph}}}]{N205}
\bibinfo{author}{\bibfnamefont{M.}~\bibnamefont{{Valluri}}},
  \bibinfo{author}{\bibfnamefont{L.}~\bibnamefont{{Ferrarese}}},
  \bibinfo{author}{\bibfnamefont{D.}~\bibnamefont{{Merritt}}},
  \bibnamefont{and} \bibinfo{author}{\bibfnamefont{C.~L.}
  \bibnamefont{{Joseph}}}, \bibinfo{journal}{\apj}
  \textbf{\bibinfo{volume}{628}}, \bibinfo{pages}{137} (\bibinfo{year}{2005}).

\bibitem[{\citenamefont{{Hernquist} et~al.}(1991)\citenamefont{{Hernquist},
  {Hut}, and {Kormendy}}}]{HHK-91}
\bibinfo{author}{\bibfnamefont{L.}~\bibnamefont{{Hernquist}}},
  \bibinfo{author}{\bibfnamefont{P.}~\bibnamefont{{Hut}}}, \bibnamefont{and}
  \bibinfo{author}{\bibfnamefont{J.}~\bibnamefont{{Kormendy}}},
  \bibinfo{journal}{\nat} \textbf{\bibinfo{volume}{354}}, \bibinfo{pages}{376}
  (\bibinfo{year}{1991}).

\bibitem[{\citenamefont{{Ferrarese}
  et~al.}(2006{\natexlab{b}})\citenamefont{{Ferrarese}, {C{\^o}t{\'e}}, {Dalla
  Bont{\`a}}, {Peng}, {Merritt}, {Jord{\'a}n}, {Blakeslee}, {Ha{\c s}egan},
  {Mei}, {Piatek} et~al.}}]{ACS-nuclei}
\bibinfo{author}{\bibfnamefont{L.}~\bibnamefont{{Ferrarese}}},
  \bibinfo{author}{\bibfnamefont{P.}~\bibnamefont{{C{\^o}t{\'e}}}},
  \bibinfo{author}{\bibfnamefont{E.}~\bibnamefont{{Dalla Bont{\`a}}}},
  \bibinfo{author}{\bibfnamefont{E.~W.} \bibnamefont{{Peng}}},
  \bibinfo{author}{\bibfnamefont{D.}~\bibnamefont{{Merritt}}},
  \bibinfo{author}{\bibfnamefont{A.}~\bibnamefont{{Jord{\'a}n}}},
  \bibinfo{author}{\bibfnamefont{J.~P.} \bibnamefont{{Blakeslee}}},
  \bibinfo{author}{\bibfnamefont{M.}~\bibnamefont{{Ha{\c s}egan}}},
  \bibinfo{author}{\bibfnamefont{S.}~\bibnamefont{{Mei}}},
  \bibinfo{author}{\bibfnamefont{S.}~\bibnamefont{{Piatek}}},
  \bibnamefont{et~al.}, \bibinfo{journal}{\apj Letters}
  \textbf{\bibinfo{volume}{644}}, \bibinfo{pages}{L21}
  (\bibinfo{year}{2006}{\natexlab{b}}).

\bibitem[{\citenamefont{{Wehner} and {Harris}}(2006)}]{Wehner-06}
\bibinfo{author}{\bibfnamefont{E.~H.} \bibnamefont{{Wehner}}} \bibnamefont{and}
  \bibinfo{author}{\bibfnamefont{W.~E.} \bibnamefont{{Harris}}},
  \bibinfo{journal}{\apj Letters} \textbf{\bibinfo{volume}{644}},
  \bibinfo{pages}{L17} (\bibinfo{year}{2006}), \eprint{astro-ph/0603801}.

\bibitem[{\citenamefont{{McLaughlin} et~al.}(2006)\citenamefont{{McLaughlin},
  {King}, and {Nayakshin}}}]{King-06}
\bibinfo{author}{\bibfnamefont{D.~E.} \bibnamefont{{McLaughlin}}},
  \bibinfo{author}{\bibfnamefont{A.~R.} \bibnamefont{{King}}},
  \bibnamefont{and}
  \bibinfo{author}{\bibfnamefont{S.}~\bibnamefont{{Nayakshin}}},
  \bibinfo{journal}{ArXiv Astrophysics e-prints}  (\bibinfo{year}{2006}),
  \eprint{astro-ph/0608521}.

\bibitem[{\citenamefont{{Kim} et~al.}(2004)\citenamefont{{Kim}, {Lee}, and
  {Spurzem}}}]{Kim-04}
\bibinfo{author}{\bibfnamefont{E.}~\bibnamefont{{Kim}}},
  \bibinfo{author}{\bibfnamefont{H.~M.} \bibnamefont{{Lee}}}, \bibnamefont{and}
  \bibinfo{author}{\bibfnamefont{R.}~\bibnamefont{{Spurzem}}},
  \bibinfo{journal}{Monthly Notices of the Royal Astronomical Society}
  \textbf{\bibinfo{volume}{351}}, \bibinfo{pages}{220} (\bibinfo{year}{2004}).

\bibitem[{\citenamefont{{Milosavljevi{\'c}}
  et~al.}(2002)\citenamefont{{Milosavljevi{\'c}}, {Merritt}, {Rest}, and {van
  den Bosch}}}]{MM-02}
\bibinfo{author}{\bibfnamefont{M.}~\bibnamefont{{Milosavljevi{\'c}}}},
  \bibinfo{author}{\bibfnamefont{D.}~\bibnamefont{{Merritt}}},
  \bibinfo{author}{\bibfnamefont{A.}~\bibnamefont{{Rest}}}, \bibnamefont{and}
  \bibinfo{author}{\bibfnamefont{F.~C.} \bibnamefont{{van den Bosch}}},
  \bibinfo{journal}{Monthly Notices of the Royal Astronomical Society}
  \textbf{\bibinfo{volume}{331}}, \bibinfo{pages}{L51} (\bibinfo{year}{2002}).

\bibitem[{\citenamefont{{Ravindranath}
  et~al.}(2002)\citenamefont{{Ravindranath}, {Ho}, and
  {Filippenko}}}]{Ravin-02}
\bibinfo{author}{\bibfnamefont{S.}~\bibnamefont{{Ravindranath}}},
  \bibinfo{author}{\bibfnamefont{L.~C.} \bibnamefont{{Ho}}}, \bibnamefont{and}
  \bibinfo{author}{\bibfnamefont{A.~V.} \bibnamefont{{Filippenko}}},
  \bibinfo{journal}{\apj} \textbf{\bibinfo{volume}{566}}, \bibinfo{pages}{801}
  (\bibinfo{year}{2002}).

\bibitem[{\citenamefont{{Graham}}(2004)}]{Graham-04}
\bibinfo{author}{\bibfnamefont{A.~W.} \bibnamefont{{Graham}}},
  \bibinfo{journal}{\apj Letters} \textbf{\bibinfo{volume}{613}},
  \bibinfo{pages}{L33} (\bibinfo{year}{2004}).

\bibitem[{\citenamefont{{Ferrarese} and {Merritt}}(2000)}]{FM-00}
\bibinfo{author}{\bibfnamefont{L.}~\bibnamefont{{Ferrarese}}} \bibnamefont{and}
  \bibinfo{author}{\bibfnamefont{D.}~\bibnamefont{{Merritt}}},
  \bibinfo{journal}{\apj Letters} \textbf{\bibinfo{volume}{539}},
  \bibinfo{pages}{L9} (\bibinfo{year}{2000}).

\bibitem[{\citenamefont{{Graham} et~al.}(2001)\citenamefont{{Graham}, {Erwin},
  {Caon}, and {Trujillo}}}]{Graham-01}
\bibinfo{author}{\bibfnamefont{A.~W.} \bibnamefont{{Graham}}},
  \bibinfo{author}{\bibfnamefont{P.}~\bibnamefont{{Erwin}}},
  \bibinfo{author}{\bibfnamefont{N.}~\bibnamefont{{Caon}}}, \bibnamefont{and}
  \bibinfo{author}{\bibfnamefont{I.}~\bibnamefont{{Trujillo}}},
  \bibinfo{journal}{\apj Letters} \textbf{\bibinfo{volume}{563}},
  \bibinfo{pages}{L11} (\bibinfo{year}{2001}).

\bibitem[{\citenamefont{{Marconi} and {Hunt}}(2003)}]{MH-03}
\bibinfo{author}{\bibfnamefont{A.}~\bibnamefont{{Marconi}}} \bibnamefont{and}
  \bibinfo{author}{\bibfnamefont{L.~K.} \bibnamefont{{Hunt}}},
  \bibinfo{journal}{\apj Letters} \textbf{\bibinfo{volume}{589}},
  \bibinfo{pages}{L21} (\bibinfo{year}{2003}).

\bibitem[{\citenamefont{{Ho}}(2004)}]{Ho-04}
\bibinfo{author}{\bibfnamefont{L.~C.~W.} \bibnamefont{{Ho}}}, in
  \emph{\bibinfo{booktitle}{Coevolution of Black Holes and Galaxies}}, edited
  by \bibinfo{editor}{\bibfnamefont{L.~C.} \bibnamefont{{Ho}}}
  (\bibinfo{year}{2004}), pp. \bibinfo{pages}{292--+}.

\bibitem[{\citenamefont{{Kaspi} et~al.}(2000)\citenamefont{{Kaspi}, {Smith},
  {Netzer}, {Maoz}, {Jannuzi}, and {Giveon}}}]{Kaspi-04}
\bibinfo{author}{\bibfnamefont{S.}~\bibnamefont{{Kaspi}}},
  \bibinfo{author}{\bibfnamefont{P.~S.} \bibnamefont{{Smith}}},
  \bibinfo{author}{\bibfnamefont{H.}~\bibnamefont{{Netzer}}},
  \bibinfo{author}{\bibfnamefont{D.}~\bibnamefont{{Maoz}}},
  \bibinfo{author}{\bibfnamefont{B.~T.} \bibnamefont{{Jannuzi}}},
  \bibnamefont{and} \bibinfo{author}{\bibfnamefont{U.}~\bibnamefont{{Giveon}}},
  \bibinfo{journal}{\apj} \textbf{\bibinfo{volume}{533}}, \bibinfo{pages}{631}
  (\bibinfo{year}{2000}).

\bibitem[{\citenamefont{{Greene} and {Ho}}(2004)}]{Greene-04}
\bibinfo{author}{\bibfnamefont{J.~E.} \bibnamefont{{Greene}}} \bibnamefont{and}
  \bibinfo{author}{\bibfnamefont{L.~C.} \bibnamefont{{Ho}}},
  \bibinfo{journal}{\apj} \textbf{\bibinfo{volume}{610}}, \bibinfo{pages}{722}
  (\bibinfo{year}{2004}).

\bibitem[{\citenamefont{{Barth} et~al.}(2004)\citenamefont{{Barth}, {Ho},
  {Rutledge}, and {Sargent}}}]{Barth-04}
\bibinfo{author}{\bibfnamefont{A.~J.} \bibnamefont{{Barth}}},
  \bibinfo{author}{\bibfnamefont{L.~C.} \bibnamefont{{Ho}}},
  \bibinfo{author}{\bibfnamefont{R.~E.} \bibnamefont{{Rutledge}}},
  \bibnamefont{and} \bibinfo{author}{\bibfnamefont{W.~L.~W.}
  \bibnamefont{{Sargent}}}, \bibinfo{journal}{\apj}
  \textbf{\bibinfo{volume}{607}}, \bibinfo{pages}{90} (\bibinfo{year}{2004}).

\bibitem[{\citenamefont{{Barth} et~al.}(2005)\citenamefont{{Barth}, {Greene},
  and {Ho}}}]{Barth-05}
\bibinfo{author}{\bibfnamefont{A.~J.} \bibnamefont{{Barth}}},
  \bibinfo{author}{\bibfnamefont{J.~E.} \bibnamefont{{Greene}}},
  \bibnamefont{and} \bibinfo{author}{\bibfnamefont{L.~C.} \bibnamefont{{Ho}}},
  \bibinfo{journal}{\apj Letters} \textbf{\bibinfo{volume}{619}},
  \bibinfo{pages}{L151} (\bibinfo{year}{2005}).

\bibitem[{\citenamefont{{Merritt} et~al.}(2001)\citenamefont{{Merritt},
  {Ferrarese}, and {Joseph}}}]{M33}
\bibinfo{author}{\bibfnamefont{D.}~\bibnamefont{{Merritt}}},
  \bibinfo{author}{\bibfnamefont{L.}~\bibnamefont{{Ferrarese}}},
  \bibnamefont{and} \bibinfo{author}{\bibfnamefont{C.~L.}
  \bibnamefont{{Joseph}}}, \bibinfo{journal}{Science}
  \textbf{\bibinfo{volume}{293}}, \bibinfo{pages}{1116} (\bibinfo{year}{2001}).

\bibitem[{\citenamefont{{Merritt} et~al.}(2004)\citenamefont{{Merritt},
  {Milosavljevi{\'c}}, {Favata}, {Hughes}, and {Holz}}}]{MMFHH-04}
\bibinfo{author}{\bibfnamefont{D.}~\bibnamefont{{Merritt}}},
  \bibinfo{author}{\bibfnamefont{M.}~\bibnamefont{{Milosavljevi{\'c}}}},
  \bibinfo{author}{\bibfnamefont{M.}~\bibnamefont{{Favata}}},
  \bibinfo{author}{\bibfnamefont{S.~A.} \bibnamefont{{Hughes}}},
  \bibnamefont{and} \bibinfo{author}{\bibfnamefont{D.~E.}
  \bibnamefont{{Holz}}}, \bibinfo{journal}{\apj Letters}
  \textbf{\bibinfo{volume}{607}}, \bibinfo{pages}{L9} (\bibinfo{year}{2004}).

\bibitem[{\citenamefont{{Mikkola} and {Valtonen}}(1992)}]{MV-92}
\bibinfo{author}{\bibfnamefont{S.}~\bibnamefont{{Mikkola}}} \bibnamefont{and}
  \bibinfo{author}{\bibfnamefont{M.~J.} \bibnamefont{{Valtonen}}},
  \bibinfo{journal}{Monthly Notices of the Royal Astronomical Society}
  \textbf{\bibinfo{volume}{259}}, \bibinfo{pages}{115} (\bibinfo{year}{1992}).

\bibitem[{\citenamefont{{Merritt} and {Milosavljevi{\'c}}}(2005)}]{Living}
\bibinfo{author}{\bibfnamefont{D.}~\bibnamefont{{Merritt}}} \bibnamefont{and}
  \bibinfo{author}{\bibfnamefont{M.}~\bibnamefont{{Milosavljevi{\'c}}}},
  \bibinfo{journal}{Living Reviews in Relativity} \textbf{\bibinfo{volume}{8}},
  \bibinfo{pages}{8} (\bibinfo{year}{2005}).

\bibitem[{\citenamefont{{Yu}}(2002)}]{Yu-02}
\bibinfo{author}{\bibfnamefont{Q.}~\bibnamefont{{Yu}}},
  \bibinfo{journal}{Monthly Notices of the Royal Astronomical Society}
  \textbf{\bibinfo{volume}{331}}, \bibinfo{pages}{935} (\bibinfo{year}{2002}).

\bibitem[{\citenamefont{{Milosavljevi{\'c}} and {Merritt}}(2003)}]{MM-03}
\bibinfo{author}{\bibfnamefont{M.}~\bibnamefont{{Milosavljevi{\'c}}}}
  \bibnamefont{and}
  \bibinfo{author}{\bibfnamefont{D.}~\bibnamefont{{Merritt}}},
  \bibinfo{journal}{\apj} \textbf{\bibinfo{volume}{596}}, \bibinfo{pages}{860}
  (\bibinfo{year}{2003}).

\bibitem[{\citenamefont{{Berczik} et~al.}(2006)\citenamefont{{Berczik},
  {Merritt}, {Spurzem}, and {Bischof}}}]{Berczik-06}
\bibinfo{author}{\bibfnamefont{P.}~\bibnamefont{{Berczik}}},
  \bibinfo{author}{\bibfnamefont{D.}~\bibnamefont{{Merritt}}},
  \bibinfo{author}{\bibfnamefont{R.}~\bibnamefont{{Spurzem}}},
  \bibnamefont{and} \bibinfo{author}{\bibfnamefont{H.-P.}
  \bibnamefont{{Bischof}}}, \bibinfo{journal}{\apj Letters}
  \textbf{\bibinfo{volume}{642}}, \bibinfo{pages}{L21} (\bibinfo{year}{2006}).

\bibitem[{\citenamefont{{Macfadyen} and {Milosavljevic}}(2006)}]{MM-06}
\bibinfo{author}{\bibfnamefont{A.~I.} \bibnamefont{{Macfadyen}}}
  \bibnamefont{and}
  \bibinfo{author}{\bibfnamefont{M.}~\bibnamefont{{Milosavljevic}}},
  \bibinfo{journal}{ArXiv Astrophysics e-prints}  (\bibinfo{year}{2006}).

\bibitem[{\citenamefont{{Rodriguez} et~al.}(2006)\citenamefont{{Rodriguez},
  {Taylor}, {Zavala}, {Peck}, {Pollack}, and {Romani}}}]{Rodriguez-06}
\bibinfo{author}{\bibfnamefont{C.}~\bibnamefont{{Rodriguez}}},
  \bibinfo{author}{\bibfnamefont{G.~B.} \bibnamefont{{Taylor}}},
  \bibinfo{author}{\bibfnamefont{R.~T.} \bibnamefont{{Zavala}}},
  \bibinfo{author}{\bibfnamefont{A.~B.} \bibnamefont{{Peck}}},
  \bibinfo{author}{\bibfnamefont{L.~K.} \bibnamefont{{Pollack}}},
  \bibnamefont{and} \bibinfo{author}{\bibfnamefont{R.~W.}
  \bibnamefont{{Romani}}}, \bibinfo{journal}{\apj}
  \textbf{\bibinfo{volume}{646}}, \bibinfo{pages}{49} (\bibinfo{year}{2006}).

\bibitem[{\citenamefont{{Navarro} et~al.}(1996)\citenamefont{{Navarro},
  {Frenk}, and {White}}}]{NFW-96}
\bibinfo{author}{\bibfnamefont{J.~F.} \bibnamefont{{Navarro}}},
  \bibinfo{author}{\bibfnamefont{C.~S.} \bibnamefont{{Frenk}}},
  \bibnamefont{and} \bibinfo{author}{\bibfnamefont{S.~D.~M.}
  \bibnamefont{{White}}}, \bibinfo{journal}{\apj}
  \textbf{\bibinfo{volume}{462}}, \bibinfo{pages}{563} (\bibinfo{year}{1996}).

\bibitem[{\citenamefont{{Moore} et~al.}(1998)\citenamefont{{Moore},
  {Governato}, {Quinn}, {Stadel}, and {Lake}}}]{Moore-98}
\bibinfo{author}{\bibfnamefont{B.}~\bibnamefont{{Moore}}},
  \bibinfo{author}{\bibfnamefont{F.}~\bibnamefont{{Governato}}},
  \bibinfo{author}{\bibfnamefont{T.}~\bibnamefont{{Quinn}}},
  \bibinfo{author}{\bibfnamefont{J.}~\bibnamefont{{Stadel}}}, \bibnamefont{and}
  \bibinfo{author}{\bibfnamefont{G.}~\bibnamefont{{Lake}}},
  \bibinfo{journal}{\apj Letters} \textbf{\bibinfo{volume}{499}},
  \bibinfo{pages}{L5+} (\bibinfo{year}{1998}).

\bibitem[{\citenamefont{{Prada} et~al.}(2004)\citenamefont{{Prada}, {Klypin},
  {Flix}, {Mart{\'{\i}}nez}, and {Simonneau}}}]{Prada-04}
\bibinfo{author}{\bibfnamefont{F.}~\bibnamefont{{Prada}}},
  \bibinfo{author}{\bibfnamefont{A.}~\bibnamefont{{Klypin}}},
  \bibinfo{author}{\bibfnamefont{J.}~\bibnamefont{{Flix}}},
  \bibinfo{author}{\bibfnamefont{M.}~\bibnamefont{{Mart{\'{\i}}nez}}},
  \bibnamefont{and}
  \bibinfo{author}{\bibfnamefont{E.}~\bibnamefont{{Simonneau}}},
  \bibinfo{journal}{Physical Review Letters} \textbf{\bibinfo{volume}{93}},
  \bibinfo{pages}{241301} (\bibinfo{year}{2004}).

\bibitem[{\citenamefont{{Graham}
  et~al.}(2006{\natexlab{a}})\citenamefont{{Graham}, {Merritt}, {Moore},
  {Diemand}, and {Terzic}}}]{Empirical3}
\bibinfo{author}{\bibfnamefont{A.~W.} \bibnamefont{{Graham}}},
  \bibinfo{author}{\bibfnamefont{D.}~\bibnamefont{{Merritt}}},
  \bibinfo{author}{\bibfnamefont{B.}~\bibnamefont{{Moore}}},
  \bibinfo{author}{\bibfnamefont{J.}~\bibnamefont{{Diemand}}},
  \bibnamefont{and} \bibinfo{author}{\bibfnamefont{B.}~\bibnamefont{{Terzic}}},
  \bibinfo{journal}{ArXiv Astrophysics e-prints}
  (\bibinfo{year}{2006}{\natexlab{a}}), \eprint{astro-ph/0608614}.

\bibitem[{\citenamefont{{Burkert}}(1995)}]{Burkert-95}
\bibinfo{author}{\bibfnamefont{A.}~\bibnamefont{{Burkert}}},
  \bibinfo{journal}{\apj Letters} \textbf{\bibinfo{volume}{447}},
  \bibinfo{pages}{L25+} (\bibinfo{year}{1995}).

\bibitem[{\citenamefont{{Salucci} and {Burkert}}(2000)}]{SB-00}
\bibinfo{author}{\bibfnamefont{P.}~\bibnamefont{{Salucci}}} \bibnamefont{and}
  \bibinfo{author}{\bibfnamefont{A.}~\bibnamefont{{Burkert}}},
  \bibinfo{journal}{\apj Letters} \textbf{\bibinfo{volume}{537}},
  \bibinfo{pages}{L9} (\bibinfo{year}{2000}).

\bibitem[{\citenamefont{{de Blok} and {Bosma}}(2002)}]{Blok-02}
\bibinfo{author}{\bibfnamefont{W.~J.~G.} \bibnamefont{{de Blok}}}
  \bibnamefont{and} \bibinfo{author}{\bibfnamefont{A.}~\bibnamefont{{Bosma}}},
  \bibinfo{journal}{Astronomy and Astrophysics} \textbf{\bibinfo{volume}{385}},
  \bibinfo{pages}{816} (\bibinfo{year}{2002}).

\bibitem[{\citenamefont{{Gentile} et~al.}(2005)\citenamefont{{Gentile},
  {Burkert}, {Salucci}, {Klein}, and {Walter}}}]{Gentile-05}
\bibinfo{author}{\bibfnamefont{G.}~\bibnamefont{{Gentile}}},
  \bibinfo{author}{\bibfnamefont{A.}~\bibnamefont{{Burkert}}},
  \bibinfo{author}{\bibfnamefont{P.}~\bibnamefont{{Salucci}}},
  \bibinfo{author}{\bibfnamefont{U.}~\bibnamefont{{Klein}}}, \bibnamefont{and}
  \bibinfo{author}{\bibfnamefont{F.}~\bibnamefont{{Walter}}},
  \bibinfo{journal}{\apj Letters} \textbf{\bibinfo{volume}{634}},
  \bibinfo{pages}{L145} (\bibinfo{year}{2005}).

\bibitem[{\citenamefont{{de Blok}}(2005)}]{Blok-05}
\bibinfo{author}{\bibfnamefont{W.~J.~G.} \bibnamefont{{de Blok}}},
  \bibinfo{journal}{\apj} \textbf{\bibinfo{volume}{634}}, \bibinfo{pages}{227}
  (\bibinfo{year}{2005}).

\bibitem[{\citenamefont{{Spekkens} et~al.}(2005)\citenamefont{{Spekkens},
  {Giovanelli}, and {Haynes}}}]{SGH-05}
\bibinfo{author}{\bibfnamefont{K.}~\bibnamefont{{Spekkens}}},
  \bibinfo{author}{\bibfnamefont{R.}~\bibnamefont{{Giovanelli}}},
  \bibnamefont{and} \bibinfo{author}{\bibfnamefont{M.~P.}
  \bibnamefont{{Haynes}}}, \bibinfo{journal}{Astronomical Journal}
  \textbf{\bibinfo{volume}{129}}, \bibinfo{pages}{2119} (\bibinfo{year}{2005}).

\bibitem[{\citenamefont{{Simon} et~al.}(2005)\citenamefont{{Simon}, {Bolatto},
  {Leroy}, {Blitz}, and {Gates}}}]{Simon-05}
\bibinfo{author}{\bibfnamefont{J.~D.} \bibnamefont{{Simon}}},
  \bibinfo{author}{\bibfnamefont{A.~D.} \bibnamefont{{Bolatto}}},
  \bibinfo{author}{\bibfnamefont{A.}~\bibnamefont{{Leroy}}},
  \bibinfo{author}{\bibfnamefont{L.}~\bibnamefont{{Blitz}}}, \bibnamefont{and}
  \bibinfo{author}{\bibfnamefont{E.~L.} \bibnamefont{{Gates}}},
  \bibinfo{journal}{\apj} \textbf{\bibinfo{volume}{621}}, \bibinfo{pages}{757}
  (\bibinfo{year}{2005}).

\bibitem[{\citenamefont{{Valenzuela} et~al.}(2005)\citenamefont{{Valenzuela},
  {Rhee}, {Klypin}, {Governato}, {Stinson}, {Quinn}, and
  {Wadsley}}}]{valenzuela-05}
\bibinfo{author}{\bibfnamefont{O.}~\bibnamefont{{Valenzuela}}},
  \bibinfo{author}{\bibfnamefont{G.}~\bibnamefont{{Rhee}}},
  \bibinfo{author}{\bibfnamefont{A.}~\bibnamefont{{Klypin}}},
  \bibinfo{author}{\bibfnamefont{F.}~\bibnamefont{{Governato}}},
  \bibinfo{author}{\bibfnamefont{G.}~\bibnamefont{{Stinson}}},
  \bibinfo{author}{\bibfnamefont{T.}~\bibnamefont{{Quinn}}}, \bibnamefont{and}
  \bibinfo{author}{\bibfnamefont{J.}~\bibnamefont{{Wadsley}}},
  \bibinfo{journal}{ArXiv Astrophysics e-prints}  (\bibinfo{year}{2005}),
  \eprint{astro-ph/0509644}.

\bibitem[{\citenamefont{{de Blok}}(2004)}]{Blok-04}
\bibinfo{author}{\bibfnamefont{W.~J.~G.} \bibnamefont{{de Blok}}}, in
  \emph{\bibinfo{booktitle}{IAU Symposium}}, edited by
  \bibinfo{editor}{\bibfnamefont{S.}~\bibnamefont{{Ryder}}},
  \bibinfo{editor}{\bibfnamefont{D.}~\bibnamefont{{Pisano}}},
  \bibinfo{editor}{\bibfnamefont{M.}~\bibnamefont{{Walker}}}, \bibnamefont{and}
  \bibinfo{editor}{\bibfnamefont{K.}~\bibnamefont{{Freeman}}}
  (\bibinfo{year}{2004}), pp. \bibinfo{pages}{69--+}.

\bibitem[{\citenamefont{{Tasitsiomi}}(2003)}]{Tasitsiomi-03}
\bibinfo{author}{\bibfnamefont{A.}~\bibnamefont{{Tasitsiomi}}},
  \bibinfo{journal}{International Journal of Modern Physics D}
  \textbf{\bibinfo{volume}{12}}, \bibinfo{pages}{1157} (\bibinfo{year}{2003}).

\bibitem[{\citenamefont{{Prugniel} and {Simien}}(1997)}]{PS-87}
\bibinfo{author}{\bibfnamefont{P.}~\bibnamefont{{Prugniel}}} \bibnamefont{and}
  \bibinfo{author}{\bibfnamefont{F.}~\bibnamefont{{Simien}}},
  \bibinfo{journal}{Astron. Astrophys.} \textbf{\bibinfo{volume}{321}},
  \bibinfo{pages}{111} (\bibinfo{year}{1997}).

\bibitem[{\citenamefont{{Navarro} et~al.}(2004)\citenamefont{{Navarro},
  {Hayashi}, {Power}, {Jenkins}, {Frenk}, {White}, {Springel}, {Stadel}, and
  {Quinn}}}]{Navarro-04}
\bibinfo{author}{\bibfnamefont{J.~F.} \bibnamefont{{Navarro}}},
  \bibinfo{author}{\bibfnamefont{E.}~\bibnamefont{{Hayashi}}},
  \bibinfo{author}{\bibfnamefont{C.}~\bibnamefont{{Power}}},
  \bibinfo{author}{\bibfnamefont{A.~R.} \bibnamefont{{Jenkins}}},
  \bibinfo{author}{\bibfnamefont{C.~S.} \bibnamefont{{Frenk}}},
  \bibinfo{author}{\bibfnamefont{S.~D.~M.} \bibnamefont{{White}}},
  \bibinfo{author}{\bibfnamefont{V.}~\bibnamefont{{Springel}}},
  \bibinfo{author}{\bibfnamefont{J.}~\bibnamefont{{Stadel}}}, \bibnamefont{and}
  \bibinfo{author}{\bibfnamefont{T.~R.} \bibnamefont{{Quinn}}},
  \bibinfo{journal}{Mon. Not. R. Astron. Soc.} \textbf{\bibinfo{volume}{349}},
  \bibinfo{pages}{1039} (\bibinfo{year}{2004}).

\bibitem[{\citenamefont{{Merritt}
  et~al.}(2005{\natexlab{a}})\citenamefont{{Merritt}, {Navarro}, {Ludlow}, and
  {Jenkins}}}]{MNLJ-05}
\bibinfo{author}{\bibfnamefont{D.}~\bibnamefont{{Merritt}}},
  \bibinfo{author}{\bibfnamefont{J.~F.} \bibnamefont{{Navarro}}},
  \bibinfo{author}{\bibfnamefont{A.}~\bibnamefont{{Ludlow}}}, \bibnamefont{and}
  \bibinfo{author}{\bibfnamefont{A.}~\bibnamefont{{Jenkins}}},
  \bibinfo{journal}{\apj Letters} \textbf{\bibinfo{volume}{624}},
  \bibinfo{pages}{L85} (\bibinfo{year}{2005}{\natexlab{a}}).

\bibitem[{\citenamefont{{Merritt}
  et~al.}(2005{\natexlab{b}})\citenamefont{{Merritt}, {Graham}, {Moore},
  {Diemand}, and {Terzic}}}]{Empirical1}
\bibinfo{author}{\bibfnamefont{D.}~\bibnamefont{{Merritt}}},
  \bibinfo{author}{\bibfnamefont{A.~W.} \bibnamefont{{Graham}}},
  \bibinfo{author}{\bibfnamefont{B.}~\bibnamefont{{Moore}}},
  \bibinfo{author}{\bibfnamefont{J.}~\bibnamefont{{Diemand}}},
  \bibnamefont{and} \bibinfo{author}{\bibfnamefont{B.}~\bibnamefont{{Terzic}}},
  \bibinfo{journal}{ArXiv Astrophysics e-prints}
  (\bibinfo{year}{2005}{\natexlab{b}}), \eprint{astro-ph/0509417}.

\bibitem[{\citenamefont{{Graham}
  et~al.}(2006{\natexlab{b}})\citenamefont{{Graham}, {Merritt}, {Moore},
  {Diemand}, and {Terzic}}}]{Empirical2}
\bibinfo{author}{\bibfnamefont{A.~W.} \bibnamefont{{Graham}}},
  \bibinfo{author}{\bibfnamefont{D.}~\bibnamefont{{Merritt}}},
  \bibinfo{author}{\bibfnamefont{B.}~\bibnamefont{{Moore}}},
  \bibinfo{author}{\bibfnamefont{J.}~\bibnamefont{{Diemand}}},
  \bibnamefont{and} \bibinfo{author}{\bibfnamefont{B.}~\bibnamefont{{Terzic}}},
  \bibinfo{journal}{ArXiv Astrophysics e-prints}
  (\bibinfo{year}{2006}{\natexlab{b}}), \eprint{astro-ph/0608613}.

\bibitem[{\citenamefont{{Ferrarese}}(2002)}]{Ferrarese-02}
\bibinfo{author}{\bibfnamefont{L.}~\bibnamefont{{Ferrarese}}},
  \bibinfo{journal}{\apj} \textbf{\bibinfo{volume}{578}}, \bibinfo{pages}{90}
  (\bibinfo{year}{2002}), \eprint{astro-ph/0203469}.

\bibitem[{\citenamefont{{Koushiappas} et~al.}(2004)\citenamefont{{Koushiappas},
  {Bullock}, and {Dekel}}}]{KBD-04}
\bibinfo{author}{\bibfnamefont{S.~M.} \bibnamefont{{Koushiappas}}},
  \bibinfo{author}{\bibfnamefont{J.~S.} \bibnamefont{{Bullock}}},
  \bibnamefont{and} \bibinfo{author}{\bibfnamefont{A.}~\bibnamefont{{Dekel}}},
  \bibinfo{journal}{MNRAS} \textbf{\bibinfo{volume}{354}}, \bibinfo{pages}{292}
  (\bibinfo{year}{2004}), \eprint{astro-ph/0311487}.

\bibitem[{\citenamefont{Bertone et~al.}(2005)\citenamefont{Bertone, Zentner,
  and Silk}}]{Bertone:2005xz}
\bibinfo{author}{\bibfnamefont{G.}~\bibnamefont{Bertone}},
  \bibinfo{author}{\bibfnamefont{A.~R.} \bibnamefont{Zentner}},
  \bibnamefont{and} \bibinfo{author}{\bibfnamefont{J.}~\bibnamefont{Silk}},
  \bibinfo{journal}{Phys. Rev.} \textbf{\bibinfo{volume}{D72}},
  \bibinfo{pages}{103517} (\bibinfo{year}{2005}), \eprint{astro-ph/0509565}.

\bibitem[{\citenamefont{Bertone}(2006)}]{Bertone:2006nq}
\bibinfo{author}{\bibfnamefont{G.}~\bibnamefont{Bertone}},
  \bibinfo{journal}{Phys. Rev.} \textbf{\bibinfo{volume}{D73}},
  \bibinfo{pages}{103519} (\bibinfo{year}{2006}), \eprint{astro-ph/0603148}.

\bibitem[{\citenamefont{{Shapiro}}(1977)}]{Shapiro-77}
\bibinfo{author}{\bibfnamefont{S.~L.} \bibnamefont{{Shapiro}}},
  \bibinfo{journal}{\apj} \textbf{\bibinfo{volume}{217}}, \bibinfo{pages}{281}
  (\bibinfo{year}{1977}).

\bibitem[{\citenamefont{Hall and Gondolo}(2006)}]{Hall:2006na}
\bibinfo{author}{\bibfnamefont{J.}~\bibnamefont{Hall}} \bibnamefont{and}
  \bibinfo{author}{\bibfnamefont{P.}~\bibnamefont{Gondolo}},
  \bibinfo{journal}{Phys. Rev.} \textbf{\bibinfo{volume}{D74}},
  \bibinfo{pages}{063511} (\bibinfo{year}{2006}), \eprint{astro-ph/0602400}.

\bibitem[{\citenamefont{{Salati} and {Silk}}(1989)}]{Salati89}
\bibinfo{author}{\bibfnamefont{P.}~\bibnamefont{{Salati}}} \bibnamefont{and}
  \bibinfo{author}{\bibfnamefont{J.}~\bibnamefont{{Silk}}},
  \bibinfo{journal}{\apj} \textbf{\bibinfo{volume}{338}}, \bibinfo{pages}{24}
  (\bibinfo{year}{1989}).

\bibitem[{\citenamefont{Moskalenko and Wai}(2006)}]{Moskalenko:2006mk}
\bibinfo{author}{\bibfnamefont{I.~V.} \bibnamefont{Moskalenko}}
  \bibnamefont{and} \bibinfo{author}{\bibfnamefont{L.}~\bibnamefont{Wai}}
  (\bibinfo{year}{2006}), \eprint{astro-ph/0608535}.

\end{thebibliography}

\end{document}